%% file: m1200.tex
\documentclass{ecai} 

\newif\ifarxiv
\arxivtrue
\newif\ifshortp
\shortpfalse

\newcommand{\shortp}[1]{
\ifshortp
\noindent\textbf{#1}
\else
\paragraph{#1.}
\fi
}

\usepackage{hyperref}
\usepackage{caption}
\makeatletter
\DeclareCaptionFormat{myformat}{%
  \sbox\@tempboxa{{\captsize{\bfseries #1.}\quad #2#3\normalsize}}%
  \ifdim \wd\@tempboxa >\hsize
    {\captsize{\bfseries #1.}\quad #2#3\normalsize}\par%
  \else
    \global \@minipagefalse
    \hbox to\hsize{\hfil\box\@tempboxa\hfil}%
  \fi
}

\makeatother

\usepackage{latexsym}
\usepackage{amssymb}
\usepackage{amsmath}
\usepackage{amsthm}
\usepackage{booktabs}
\usepackage{enumitem}
\usepackage{graphicx}
\usepackage{color}

\newcommand{\BibTeX}{B\kern-.05em{\sc i\kern-.025em b}\kern-.08em\TeX}

\usepackage{xcolor}
\usepackage{listings}
\usepackage{mdframed}

\definecolor{template}{RGB}{72,67,73}
\definecolor{boxgray}{RGB}{247,240,240}
\definecolor{border}{RGB}{	42, 39, 42}
\definecolor{input}{RGB}{109,163 77}
\definecolor{llm}{RGB}{0,167,225}
\definecolor{sampled}{RGB}{	246, 126, 125}

\lstdefinestyle{interaction}{
  backgroundcolor=\color{boxgray}, % set the background color
  frame=single, % frame code snippets
  captionpos=b,
  framerule=1pt, % thickness of frame
  rulecolor=\color{border}, % frame color
  basicstyle=\rmfamily\small\color{template},
  columns=fullflexible,
  keepspaces=true,
  breaklines=true,   % Enables text wrapping
  breakindent=0pt,  % No indentation for wrapped lines
  moredelim=[is][\color{template}]{<template>}{</template>},
  moredelim=[is][\color{template}]{<input>}{</input>},
  moredelim=[is][\color{llm}]{<llm>}{</llm>},
moredelim=[is][\color{sampled}\textsc]{<sampled>}{</sampled>},
}

\newmdenv[backgroundcolor=boxgray,linewidth=0pt,innerleftmargin=5pt,innerrightmargin=5pt,innertopmargin=5pt,innerbottommargin=5pt]{graybox}

\usepackage{booktabs} %layout
\usepackage{array} %center
\usepackage{multirow} % rows 
\usepackage{multicol}

\usepackage{placeins}
\usepackage{stfloats}
\usepackage{amsmath}

\usepackage{cleveref} % citing
\Crefname{lstlisting}{listing}{Listings}
\Crefname{lstlisting}{Listing}{Listings}

\begin{document}

%%%%%%%%%%%%%%%%%%%%%%%%%%%%%%%%%%%%%%%%%%%%%%%%%%%%%%%%%%%%%%%%%%%%%%%%

\begin{frontmatter}

%%% Use this command to specify your submission number.
%%% In doubleblind mode, it will be printed on the first page.

\paperid{1200} 

%%% Use this command to specify the title of your paper.

\title{SUBER: An RL Environment with Simulated\\Human Behavior for Recommender Systems}

%%% Use this combinations of commands to specify all authors of your 
%%% paper. Use \fnms{} and \snm{} to indicate everyone's first names 
%%% and surname. This will help the publisher with indexing the 
%%% proceedings. Please use a reasonable approximation in case your 
%%% name does not neatly split into "first names" and "surname".
%%% Specifying your ORCID digital identifier is optional. 
%%% Use the \thanks{} command to indicate one or more corresponding 
%%% authors and their email address(es). If so desired, you can specify
%%% author contributions using the \footnote{} command.

\author[A]{\fnms{Nathan}~\snm{Corecco}\footnote{Equal contribution.}}
\author[A]{\fnms{Giorgio}~\snm{Piatti}\footnotemark}
\author[A]{\fnms{Luca A.}~\snm{Lanzendörfer}\thanks{Corresponding Author. Email: lanzendoerfer@ethz.ch}} 
\author[B,C]{\fnms{Flint Xiaofeng}~\snm{Fan}} 
\author[A]{\fnms{Roger}~\snm{Wattenhofer}} 

\address[A]{ETH Zurich}
\address[B]{National University of Singapore}
\address[C]{A*STAR Singapore}

%%% Use this environment to include an abstract of your paper.

\begin{abstract}
    Reinforcement learning (RL) has gained popularity in the realm of recommender systems due to its ability to optimize long-term rewards and guide users in discovering relevant content.
    However, the successful implementation of RL in recommender systems is challenging because of several factors, including the limited availability of online data for training on-policy methods. This scarcity requires expensive human interaction for online model training. Furthermore, the development of effective evaluation frameworks that accurately reflect the quality of models remains a fundamental challenge in recommender systems. To address these challenges, we propose a comprehensive framework for synthetic environments that simulate human behavior by harnessing the capabilities of large language models (LLMs). We complement our framework with in-depth ablation studies and demonstrate its effectiveness with experiments on movie and book recommendations. Using LLMs as synthetic users, this work introduces a modular and novel framework to train RL-based recommender systems. The software, including the RL environment, is publicly available on \url{https://github.com/SUBER-Team/SUBER}.
    % The software, including the RL environment, is been uploaded to the submission system, and will be open-sourced upon acceptance.
\end{abstract}

\end{frontmatter}

\section{Introduction}
In an age defined by the ubiquitous presence of digital platforms in both leisure and commerce, recommender systems have emerged as instrumental tools in guiding user choices. From Netflix tailoring movie suggestions to match the cinematic tastes of users to Amazon presenting personalized products lists to shoppers, recommendation systems are the engines driving enhanced user experiences and the engagement of the platform \citep{steck2021deep,agrawal2023enhancing}.

Reinforcement Learning (RL), with its principles rooted in learning by interaction, provides a compelling approach to dynamically and adaptively tailor recommendations. Recommender systems should take into account both short- and long-term rewards and direct the interests of users towards appropriate recommendations. An increasing body of research has investigated the use of RL in recommender systems \citep{ie2019reinforcement,chen2019top,liu2022monolith,afsar2022reinforcement,lin2023survey}. Although promising, the use of RL for recommendation systems comes with its own set of challenges:

\emph{Data Availability}:  RL algorithms require a significant amount of data from interactions with the environment to learn effective policies. However, in the case of recommender systems, users may quickly abandon the service if they receive random or irrelevant recommendations. This makes it impractical to collect the large amount of data needed to train an RL model without compromising the user experience \citep{zhang2016collective}. 

\emph{Unknown user model}: In RL, a reward function is crucial to allow the model to learn effectively. In the context of recommender systems, designing an appropriate synthetic reward function that accurately reflects user satisfaction or preferences can be challenging due to the complexity of modeling human behavior \citep{chen2019generative,shi2019virtual}.

\emph{Model evaluation}: A key challenge in recommender systems is the evaluation of models without directly interacting with real users, thus avoiding any potential negative impact on the user experience. On the other hand, evaluating on offline data does not guarantee good recommendation performance in the real world \citep{shani2011evaluating,garcin2014offline}.

In this work, we propose a "Simulated User Behavior for Recommender Systems" (SUBER), a novel framework for recommender systems to address the aforementioned challenges. SUBER is a framework for synthetic environments that use Large Language Models (LLM) at its core. SUBER leverages recent advances in LLMs to simulate human behavior~\citep{park2023generative,argyle2023out}. Furthermore, by training on large amounts of data, LLMs have obtained inherent knowledge about movies, books, and various other objects. These strengths, the ability to mimick human behavior coupled with vast knowledge about humanity, uniquely position LLMs as a powerful tool to simulate users in synthetic environments for recommender systems. Therefore, SUBER serves as a versatile playground for researchers, allowing them to experiment with different LLM configurations, fine-tune user specifications, and improve their RL strategies. Our contributions can be summarized as follows:
\vspace{1em}
\begin{itemize}
[itemsep=0.7em,topsep=-0.25em
,leftmargin=1.6em
]
    \item We introduce SUBER, a versatile framework for training and evaluating RL-based recommender systems. Our framework includes a gym environment with an LLM designed to simulate human behavior and rate recommended items accordingly.
    \item We conduct extensive ablation studies to assess the impact of each component in our framework.
    Moreover, we present findings across multiple LLM families, revealing their influence on the environment's performance and highlighting their effectiveness in replicating human behavior for item recommendations.   
    \item We experimentally validate our environment using both movie and book recommendation settings. Additionally, we have made all code available as open-source.
\end{itemize}

%%%%%%%%%%%%%%%%%%%%%%%%%%%%%%%%%%%%%%%%%%%%%%%%%%%%%%%%%%%%

%%%%%%%%%%%%%%%%%%%%%%%%%%%%%%%%%%%%%%%%%%%%%%%%%%%%%%%%%%%%
\section{Related Work}
\label{sec:related_work}

\begin{table*}[ht!]
\centering
\caption{Comparison of simulation environments for recommender systems. We list whether the user and item datasets are real or synthetic. Simulation Engine indicates the different approaches used. For the evaluation strategy, we distinguish between offline evaluation in the original dataset used to train the simulator, online testing on a platform, sanity checks, and case studies.
}
\label{table:comparison_simulator}
\begin{tabular}
{m{0.25\linewidth}m{0.08\linewidth}m{0.08\linewidth}m{0.21\linewidth}m{0.23\linewidth}}
\toprule
Simulators & User dataset & Item dataset & Simulation engine & Evaluation strategy \\
\midrule
Adversarial  \citep{chen2019generative} & Real & Real & GAN & Offline  \\
VirtualTaobao \citep{shi2019virtual} & Real & Real & GAN  & Online  \\
RL4RS \citep{wang2023rl4rs}  & Real & Real &   Transformer &  Online\\
KuaiSim  \citep{zhao2023kuaisim} & Real & Real & Transformer & Offline \\ 
RecoGym   \citep{rohde2018recogym} & Synthetic & Synthetic  & Statistical modelling &  Sanity checks \\
RecoSim  \citep{ie2019recsim} & Synthetic & Synthetic & Statistical modelling & Case studies \\
SUBER  (our) &  Synthetic & Real & LLM & Sanity checks \& case studies  \\
\bottomrule
\end{tabular}

\end{table*}

\shortp{RL for Recommender Systems}
Platforms such as YouTube \citep{ie2019reinforcement,chen2019top} and BytePlus \citep{liu2022monolith} are two of many recent successful examples of training and evaluating recommender systems with online data. 
% In industry, researchers usually have more access to online systems: they can collect more data, train online methods on a subset of users, and perform A/B testing to evaluate the performance of new models.
Traditional and neural recommender systems and have been extensively researched in the past three decades \citep{goldberg1992using,su2009survey,bobadilla2013recommender,shi2014collaborative,lu2015recommender,zhang2019deep}. However, since our work focuses on RL in recommender systems (RL4Rec), we limit the related work to this area of research.
Although RL4Rec has been the subject of several studies, most of the work has been based primarily on training and evaluation based on offline datasets~\citep{afsar2022reinforcement,lin2023survey}.
% Applying RL to recommender systems (RL4Rec) has been the subject of several studies. 
% \citep{afsar2022reinforcement} and \citep{lin2023survey} investigate the RL approaches for recommending items to users. They show how a majority        of the work has relied predominantly on training and evaluation based on offline datasets. 
As indicated by \citet{afsar2022reinforcement}, online assessment is the preferred approach for evaluation. However, it presents significant challenges with respect to complexity and expense. In contrast, offline evaluation takes place in a static and biased environment. Therefore, Afsar et al. call for the creation of a versatile simulator for RL4Rec similar in nature to OpenAI's Gym for conventional RL tasks~\citep{brockman2016openai}. Additional challenges exist in the wider domain of RL, specifically regarding issues related to off-policy learning and offline policy evaluation, which become even more complex when incorporated into recommender systems \citep{precup2001off,gelada2019off,kumar2019stabilizing}.

Notable efforts have been made to address the limitations of offline learning in recommender systems.
To this end, many simulation environments for recommender systems have been developed. \citet{rohde2018recogym} presented RecoGym, a synthetic environment that addresses exploding variance by simulating user responses to different recommendation strategies. RecSim \citep{ie2019recsim} is a customizable synthetic simulation platform that incorporates various assumptions about user preferences, item familiarity, user latent states and dynamics, and choice models. \citet{chen2019generative} proposed a generator that captures the underlying distribution of historical user interactions and learns to generate realistic interactions. Extending this idea, \citet{shi2019virtual} proposed VirtualTaobao, a virtual shopping environment, and demonstrated the superiority of policies developed in this framework over traditional supervised techniques in real-world settings.
\citet{wang2023rl4rs} introduced the RL4RS dataset to address the lack of validated simulation environments and advanced evaluation methods in RL-based recommender system research. The dataset is collected from a NetEase game and anonymized through a three-step process.
\citet{zhao2023kuaisim} propose KuaiSim, a versatile environment that provides user feedback with multi-behavior and cross-session responses, supporting three tasks: request-level list-wise recommendation task, whole-session-level sequential recommendation task, and cross-session-level retention optimization task.
Unlike previous approaches, our work leverages natural language by using LLMs to simulate user behavior. In addition, our framework is not dataset dependent, and therefore, the set of users and items are not restricted to specific domains.
%This flexibility allows for unique environmental behaviors, such as incorporating new items.

% Notable efforts have been made to address the limitations of offline learning in recommender systems. \citep{yu2017seqgan} introduced SeqGAN, which combines Generative Adversarial Networks (GANs) and an RL-inspired generator to generate sequences. The discriminator provides a reward signal at the end of the sequence. In contrast, \citep{bai2019model} proposed a two-component generator consisting of a recommendation agent and a user behavior model. Their approach, which goes beyond simple sequence generation, uses adversarial training to address biases in the user model and subsequently reduces variance during training.

% GANs have also been studied as a means to simulate the dynamics of user behavior. \citep{chen2019generative} and \citep{zhao2019toward} proposed a generator that captures the underlying distribution of historical interactions of users and learns to generate realistic interactions. Extending this idea,~\citep{shi2019virtual} proposed a virtual shopping environment and demonstrated the superiority of policies developed in this space over traditional supervised techniques in real-world settings. GANs trained on offline data primarily generate data samples that mimic patterns from their training sets~\citep{huang2020keeping}. However, this intrinsic design can make it difficult to accurately simulate user behavior in scenarios with novel or unobserved content. Therefore, we discuss how LLMs can be better suited to simulate human behavior.

\pagebreak
\shortp{Large Language Models} 
There have been significant recent advances in the field of LLMs. These models are primarily based on the transformer architectures introduced by \citet{vaswani2017attention} and have continued to grow in size, capability, and performance. The Generative Pre-trained Transformer (GPT) series by OpenAI \citep{brown2020language,openai2023gpt4} is one of the most notable developments in this area, demonstrating the immense potential and scalability of transformer-based models. The recent release of foundation language models such as Llama-1 and Llama-2 \citep{touvron2023llama1,touvron2023llama2}, has democratized the access to these large LLMs. This has paved the way for the creation of instruction-following models such as Vicuna \citep{zheng2023judging} and Mistral \citep{jiang2023mistral}. Meanwhile, numerous efforts have focused on optimizing the memory consumption and inference speed of LLMs. For example, GPTQ \citet{frantar2022gptq} compressed the model parameters to 4 bits, allowing larger models to run on hardware with less memory and without significant loss of performance.

LLMs can generate textual content that rivals the quality of human-generated text \citep{brown2020language}. However, their applications go beyond text generation. \citet{park2023generative} demonstrated how LLMs can be used to simulate human behavior. These simulated agents wake up, cook, go to work, make decisions, and reflect on past experiences in a believable manner. Furthermore, \citet{argyle2023out} suggests using language models as surrogates for certain demographic groups within social science research. Their study demonstrates how conditioning GPT-3 on the socio-demographic backgrounds of real human subjects can accurately replicate response distributions among diverse human subgroups.

Contemporary work has also integrated LLMs into recommender systems. \citet{kang2023llms} demonstrated that fine-tuned LLMs outperform traditional supervised methods in predicting user ratings with less training data, while \citet{wang2023recmind} employed LLMs as a recommendation agent, showcasing their potential to improve recommender systems. Both works show how LLMs can act as a good predictor of the ratings that a user would assign to an item. The authors further investigated whether LLMs can also be used as a recommender directly; they restricted their experiment to choosing an item from a list of 100 items. However, this task is still challenging for LLMs, as they must have knowledge of the entire set of possible items to be recommended. The limited context length does not allow one to provide a list of all possible items in the prompt to an LLM.
Therefore, to date, the application of large language models (LLMs) as recommender systems has yet to exceed the performance of traditional recommender systems, which encompass both classical supervised algorithms and those based on reinforcement learning techniques.
% Compared to these contemporary works, our main difference is the integration of LLMs as simulation environments for item recommendation, while related works train the LLM to be the recommender system itself.
Our work diverges from these approaches by leveraging LLMs as simulation environments for item recommendation, in contrast to prior efforts that focused on training LLMs to function as the recommender system itself.

%%%%%%%%%%%%%%%%%%%%%%%%%%%%%%%%%%%%%%%%%%%%%%%%%%%%%%%%%%%%
\section{Framework}
\label{sec:framework}

\begin{figure*}[t]
  \begin{center}
    \includegraphics[width=\linewidth]{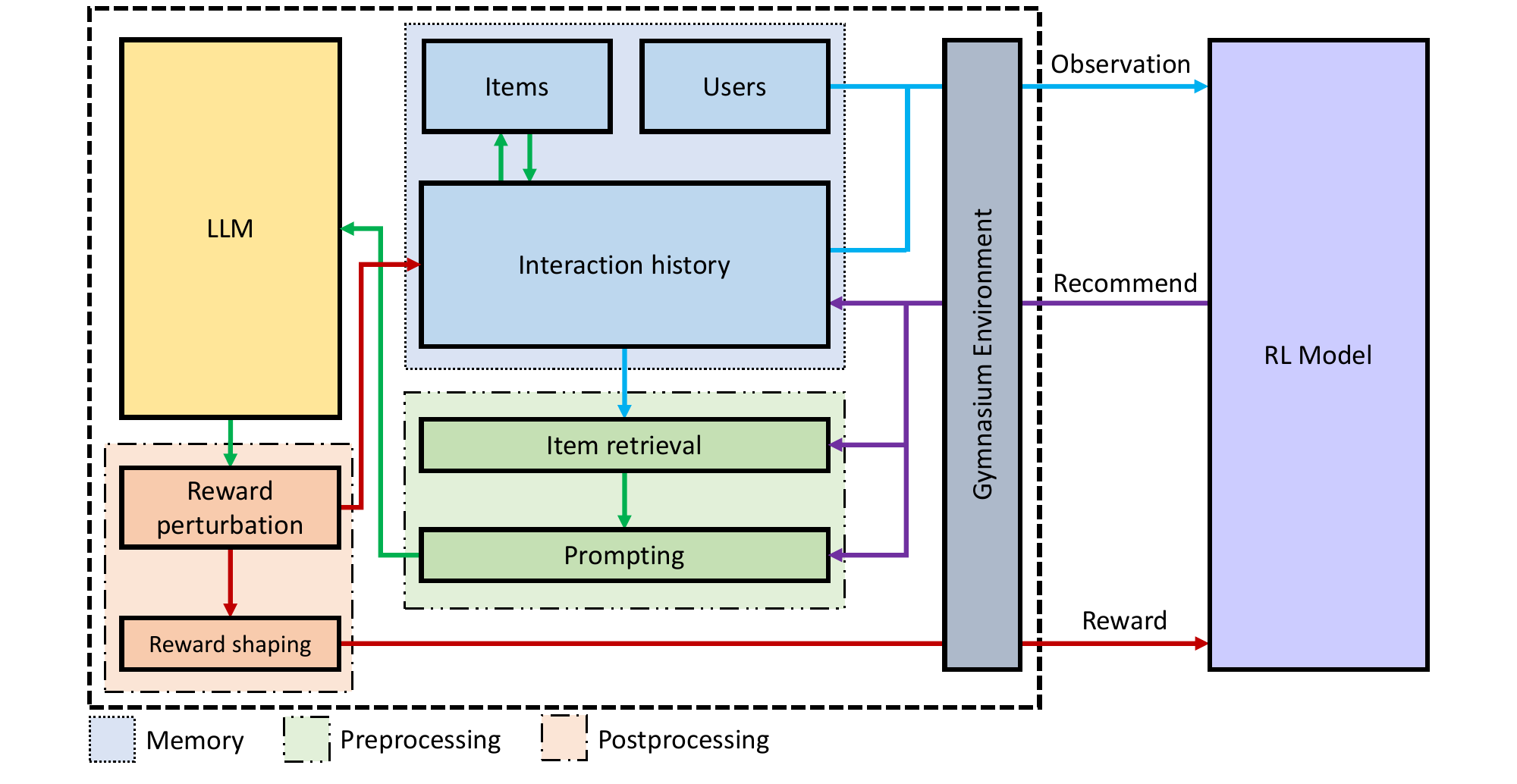}
    \caption{Overview of SUBER. The environment is built as a modular framework where each component can be modified as required. The basic control flow is as follows: The environment provides an observation using the memory module; the RL model returns an item recommendation in the form of an action, which is processed into a prompt by the memory and preprocessing component before being passed to the LLM. The score returned by the LLM is postprocessed, stored in memory and returned as a reward to the RL model.}
    \label{fig:env}
  \end{center}
\end{figure*}

To address the aforementioned challenges of data availability, unknown user model, and model evaluation, we propose SUBER, an environment designed to simulate human behavior through the integration of LLMs. SUBER serves a dual purpose by generating synthetic data and harnessing the capabilities of LLMs to replicate the behavior of individuals with unknown patterns. Additionally, this dynamic environment can serve as a model evaluation tool for recommender systems.

SUBER consists of an LLM component and three separate modules that contain multiple individual components. An overview of the overall structure is presented in \Cref{fig:env}. The internal memory module of the environment contains two separate datasets, one for users and one for items. The environment also includes a preprocessing module that retrieves raw data from the memory module and transforms it to ensure compatibility with the LLM. Finally, a post-processing component transforms the output produced by the LLM before returning it to the RL model.

The interaction with an RL model involves the following information flow: initially, the environment selects a user from memory, along with their interaction history (i.e., items and associated ratings) as the observation for the RL model. The RL model then recommends an item to the user as its action, with an action space equal to the number of items in the environment. The action and observation are subsequently processed through the preprocessing module, the LLM component, and the postprocessing module. Finally, the environment returns a reward corresponding to the post-processed rating predicted by the LLM. We describe each module in more detail in the following sections.

Our environment is designed with easy accessibility and extensibility in mind. Therefore, we chose a modular approach and based the environment interface on the Gymnasium standardized API \citep{towers_gymnasium_2023}. Different components can be modified at will, providing additional flexibility in future design choices.

\subsection{Memory}
We introduce the following notation. We define $U$ as the set of users and $I$ as the set of items. For every pair of user-items $(u,i)\in U\times I$, we have a set $R_{u,i}$ that records all interactions between user $u$ and item $i$. Similarly, for every user $u$ we define with $R_u$ the set of all interactions with all items, defined as follows:
\begin{equation}
R_u = \{(i, h) \vert i \in I, h \in R_{u,i}\}.
\end{equation}
The memory module consists of three components: an item dataset, a user dataset, and a record of all interactions between users and items. This interaction history stores the set of interactions $R_{u,i}$ for each pair of user-items $(u,i)$. Every interaction between the RL model and the environment produces a new interaction record between a user and an item, which is added to the interaction history.

\subsection{Pre-processing}
\label{subsec:preprocessing}

\shortp{Item Retrieval}
As the RL model interacts with the environment, the history of the interaction increases. It may be challenging to extract relevant information from long histories, and the increasing duration of the history will probably exceed the context length of current LLMs~\citep{park2023generative}. To address this issue, we propose an item-retrieval component responsible for retrieving the most appropriate items for the current query from the interaction history of a user.
% This process ensures that the LLM can respond authentically and consistently to queries. 
Additionally, as user interests and preferences may evolve over time, relying solely on user features may not accurately capture current interests. Therefore, historical rating data are used to provide a more detailed depiction of their evolving preferences.

\begin{figure*}[t]
  \begin{center}
    \includegraphics[width=\linewidth]{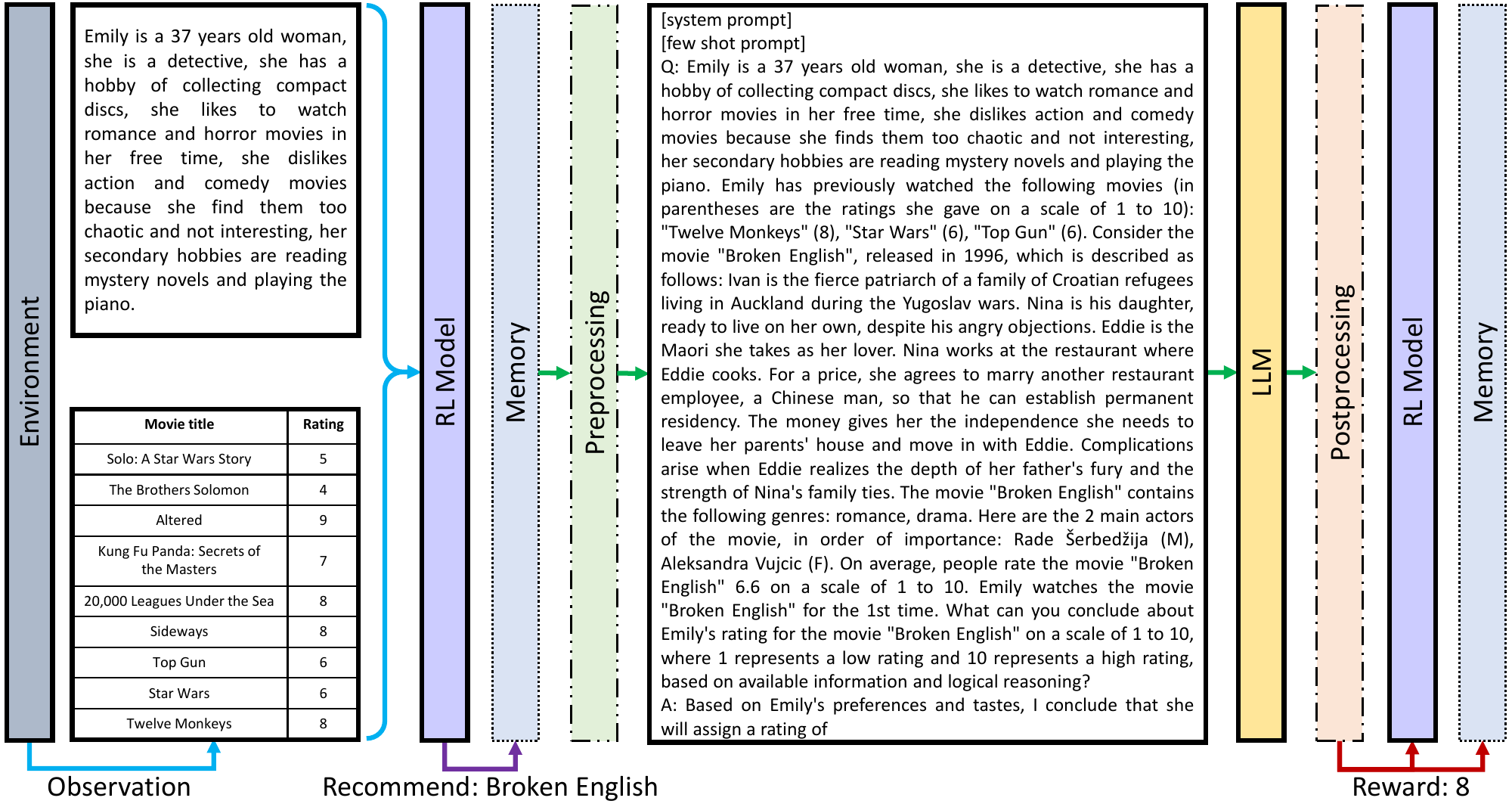}
    \vspace{2pt}
    \caption{Pipeline of one interaction between the RL model and SUBER. The environment provides an observation in the form of a user description and user-item interaction history to the RL model. The RL model then recommends an item, which is processed into a prompt together with the user description and interaction history. The LLM uses this prompt to generate a reward for the recommended item. The reward is stored as part of the user-item interaction history and returned to the RL model.}
    \label{fig:LLM_interaction}
  \end{center}
\end{figure*}

% \paragraph{Prompting}
% \label{para:prompting}
\shortp{Prompting}
The prompting component aggregates the information retrieved by the item retrieval component, creating a prompt that contains the necessary details for the LLM, including the user and query item data. The objective of this prompt is to enable the LLM to accurately predict the rating of the current query item. An example of such a prompt as part of an interaction example can be seen in \Cref{fig:LLM_interaction}.

\subsection{Postprocessing}
\label{subsec:postprocessing}

% \paragraph{Reward Perturbation}
% \label{para:reward_perturbation}

\shortp{Reward Perturbation} The reward perturbation component introduces noise into the ratings generated by the LLM. This component functions as a simulation of ``concept drift" for users \citep{vzliobaite2016overview}. Concept drift refers to the notion that users may change their interests over time and are unlikely to maintain static preferences.
% Therefore, perturbing the LLM prediction, together with historical data of the user, can impact the model's predictions.

% \paragraph{Reward shaping}
% \label{para:reward_shaping}
\shortp{Reward Shaping} Similarly to the reward perturbation component, reward shaping modifies the reward. However, unlike the perturbed reward which is added to the memory, the reward modified by the reward shaping component is returned directly to the RL model and is not stored in memory. The reward shaping module aims to reflect changes in the reward that are not related to a change in the preference of a user, such as spontaneous decisions or fleeting interests.

%%%%%%%%%%%%%%%%%%%%%%%%%%%%%%%%%%%%%%%%%%%%%%%%%%%%%%%%%%%%
\section{Experiments}  
\label{sec:experiments}

To evaluate SUBER, we followed the approach of 
\citet{rohde2018recogym} and \citet{ie2019recsim}. We perform sanity checks and case studies, which we present in \Cref{subsec:ablations} and \Cref{subsec:experiments_rl}.
To achieve this, we implemented a movie recommendation and a book recommendation environment in our framework. In the following sections, we discuss our implementation and design choices for these environments, as well as our ablation study and experiments. For the movie setting, we use rewards from 1 to 10, similar to TMDB \footnote{\url{https://www.themoviedb.org/}}, while for the book setting we use rewards from 1 to 5, as found in the Amazon Reviews Dataset \citep{ni-etal-2019-justifying}.

For both environments, we created a dataset of synthetic users using Vicuna \citep{zheng2023judging} with Guidance \citep{guidance2023}. To generate user descriptions, we condition the LLM with information such as the age, liked and disliked genres, hobbies, and profession of the user (cf.~\Cref{listing:prompt}). We generate the user age by sampling from the age distribution in the United States~\citep{age_statistic}, %while hobbies and professions were sampled from a precompiled list 

We randomly select a hobby and a profession from predefined lists (cf. \ifarxiv~\Cref{app:list}\else Appendix G in \cite{corecco2024suber}\fi). These hobby lists are divided into two categories: one tailored to children (aged 4-17) and another for adults (aged 18-75). Users not of working age are assigned the profession ``student," while those of retirement age are categorized as ``retired."
For the movie dataset we use MovieLens  (\textit{ml-latest-small})  \citep{harper2015movielens} and collect the respective movie features from TMDB. For the book dataset, we used a subset of the Amazon Book Dataset. For more details, see \ifarxiv\Cref{app:movies_env} and \Cref{app:books_env}.
\else
Appendix A and Appendix B in \cite{corecco2024suber}.
\fi

\clearpage
\begin{lstlisting}[style=interaction, label=listing:prompt, caption={User generation and characteristic assignment process example by Vicuna with guidance. {\color{template}Black text} shows the template and the instruction, \textsc{\color{sampled}red text} marks the sampled information from external distributions, {\color{llm}blue text} indicates the content generated by LLM.}]
<template>Can you generate details for a person, you need to generate a name, an age, a hobby, a job and a detailed, long and original description that contains the persons interests and secondary hobbies. Please outline the cinematic preferences of the individual, detailing their favorite and least favorite genres. Kindly provide explanations for each genre preference. </template> 
<template>Name: </template><llm>Emily Johnson</llm>, <template>Age: </template><sampled>37</sampled>, <template>Gender: </template><llm>F</llm>, <template>Hobby: </template><sampled>compact discs</sampled>, <template>Job: </template><sampled>detective</sampled>
<template>Genres liked: </template><llm>romance, horror</llm>, <template>Genres diskliked: </template><llm>action, comedy</llm>
<template>Description: she is </template><llm>a detective, she has a hobby of collecting compact discs, she likes to watch romance and horror movies in her free time, she dislikes action and comedy movies because she find them too chaotic and not interesting, her secondary hobbies are reading mystery novels and playing the piano.</llm>
\end{lstlisting}

\subsection{Setup}
\label{subsec:setup}
We implemented three different approaches for the retrieval component: feature retrieval, recency retrieval, and similarity retrieval. The feature-based approach retrieves items based on the Sorensen Coefficient of movie genres, actors, director, and average rating, while for books, we use book category, author, and average rating. The recency algorithm retrieves the most recent interactions. The similarity approach retrieves items from the history based on their similarity to the query item. We generate item-description embeddings using a Sentence-T5 model~\citep{ni2022sentence} and compute their similarities based on the cosine distance. To select the item-rating pair to retrieve from memory, we compute the similarity between the query item and all items previously viewed by the current user, selecting the items with the highest similarity.
We use the items returned from the retrieval component to construct a prompt to query the LLM. The LLM is tasked with generating a rating of the queried item by the current user, where the queried item corresponds to the item suggested by the recommender system. We construct the prompt such that the user description comes first, allowing us to leverage the key-value cache \citep{pope2023efficiently}, eliminating the need to recalculate all intermediate embeddings within the layers of the LLM for already encountered prefixes, therefore, increasing execution speed. 
Furthermore, we experimented with one-shot and two-shot prompting to improve model performance, which has been shown to increase generation quality~\citep{brown2020language}. In addition to the default system prompt , we created a custom system prompt (see \Cref{listing:our_system_prompt} for movie and 
\ifarxiv\Cref{app:books_env}\else Appendix B in \cite{corecco2024suber} \fi~for books).

\begin{lstlisting}[style=interaction, label=listing:our_system_prompt, caption={An advanced system prompt guiding the model to provide personalized and unbiased movie ratings.}]
<template>You are a highly sophisticated movie rating assistant, equipped with an advanced understanding of human behavior. Your mission is to deliver personalized movie recommendations by carefully considering the unique characteristics, tastes, and past-seen films of each individual. When presented with information about a specific movie, you will diligently analyze its plot, primary genres, actors, and average rating. Using this comprehensive understanding, your role is to provide thoughtful and accurate ratings for movies on a scale of 1 to 10, ensuring they resonate with the person's preferences and cinematic inclinations. Remain impartial and refrain from introducing any biases in your predictions. You are an impartial and reliable source of movie rating predictions for the given individual and film descriptions.</template>
\end{lstlisting}

Tokenization ambiguity can become an issue when generating numbers with LLMs. Since we are dealing with ratings on a scale from one to ten, and because the number ``10" can be tokenized in two different ways, this can cause unwanted side effects. To tackle this challenge, we tested two additional strategies for the movie setting: shifting all rewards to the scale of 0-9, and using words for numbers from ``one" to ``ten."

We experimented with various quantized versions of Llama, Vicuna, Mistral, using LLMs that could run within a 24GB memory limit. A list of the models used in our experiments can be found in \ifarxiv\Cref{app:ablations_all_extended}\else Appendix D in \cite{corecco2024suber}\fi
. All models were quantized using GPTQ. Since different LLMs influence the simulation of human behavior in different ways, it is important to highlight the inherent trade-off between model size and processing speed. In particular, during training of an RL model, a fast environment is desirable to acquire more samples in a shorter time span. However, smaller LLMs may not adequately emulate the desired human behavior of our synthetic users.

For the reward perturbation experiment, we compared Gaussian noise and greedy noise. Greedy noise alters the LLM rating by 1 with a probability of $q$, while it remains unchanged with a probability of $1-q$.

Our implementation of reward shaping operates on the following premise: as a user engages with an item more frequently, their interest in revisiting it diminishes. In contrast, as time passes, the likelihood that the user interacts with the item increases again \citep{10.1086/662996}. Given this insight, let us consider a user $u$ from the set $U$ and an item $i$ with which the user has interacted $n_{ui}$ times. When a time span of $\Delta t$ has passed since the last interaction with the item, the reward $r$ undergoes a reshaping process, characterized by the following equation:
\begin{equation}
\label{eqn:reshaping}
r \gets \max(1, \lfloor{r\cdot q^{n_{ui}/\Delta t}}\rfloor),
\end{equation}
where $q\in [0,1]$. This adjustment takes into account both the frequency of user interaction with the item and the time elapsed since their last interaction, resulting in the modified reward $r$.

\subsection{Ablations}
\label{subsec:ablations}

To determine the effect of each component in our environment, we performed ablations across four different test cases. In this section, we present the high-level idea; for more details, see \ifarxiv\Cref{app:experiment_details}\else Appendix C in the supplementary material \cite{corecco2024suber}\fi.

% \comment{Short sentence on default config citing \Cref{table:defaultEnvConfig}}
% We discuss each of the four test cases in the following sections. 

% \begin{table}[h]
% \caption{}
% \centering
% \label{table:defaultEnvConfig}
% \begin{tabular}{ll}
% \toprule
% \textbf{Component} & \textbf{Default config} \\
% \midrule
% LLM & Vicuna-v1.5 13B  \\
% User dataset & detailed \\
% Retrieval & most similar (3)  \\
% Prompting & \\
% Reward perturbator & \\
% Reward shaping &  \\
% \bottomrule
% \end{tabular}
% \end{table}

 \shortp{Genres/Categories} 
 We assess the environment's ability to recognize movie and book genres and its ability to correlate those genres with user preferences to accurately predict ratings. User profiles were manually created for each movie genre, ensuring that they expressed a preference for the selected genre while disliking all others. Afterwards, we queried the environment with users and movies from both their favored and disliked genres. The accuracy of rating predictions is used to measure performance. A similar process is used for the book environment, replacing movie genres with book categories.

\begin{table*}[th]
\centering
\caption{Ablation results for the movie setting using \textit{Mistral 7B} as our environment. We test the LLM on coherency and realistic ratings for user-movie interactions. We achieve best performance with 0-9 digit rating scale, 2-shot prompting, and our custom system prompt.}

\label{table:ablations_movies_prompt}
\begin{tabular}{
m{0.08\linewidth}
m{0.07\linewidth}
m{0.07\linewidth}|
m{0.09\linewidth}
m{0.13\linewidth}
m{0.11\textwidth}
m{0.10\textwidth}
m{0.11\textwidth}}
\toprule
\multicolumn{3}{c}{Prompt component} \\
Rating scale  & N-shot & System prompt & Genres $\uparrow$ & High/Low $\uparrow$  & Collection \newline of movies $\uparrow$ & Similarity \newline to ML $\uparrow$ & Agg. score $\uparrow$  \\
\midrule
0-9 & 0-shot & default & 0.80$\pm$0.00& \textbf{1.00$\pm$0.00}& 0.67$\pm$0.02 & 0.54$\pm$0.00 & 0.75$\pm$0.01 \\

0-9 & 0-shot& custom & \textbf{0.87$\pm$0.00}& \textbf{1.00$\pm$0.00}& 0.68$\pm$0.02 & 0.70$\pm$0.00 & 0.81$\pm$0.01 \\

0-9 & 1-shot & default & 0.72$\pm$0.00 & 0.96$\pm$0.00 & \textbf{0.71$\pm$0.03} & 0.73$\pm$0.01 & 0.78$\pm$0.01 \\
0-9 & 1-shot & custom  & 0.81$\pm$0.00& \textbf{1.00$\pm$0.00}& \textbf{0.71$\pm$0.02} & 0.78$\pm$0.00 & \textbf{0.82$\pm$0.01} \\
0-9 & 2-shot & default & 0.78$\pm$0.00 & 0.99$\pm$0.00 & 0.69$\pm$0.01 & \textbf{0.80$\pm$0.00} & \textbf{0.82$\pm$0.00} \\
0-9 & 2-shot  & custom  & 0.79$\pm$0.00& \textbf{1.00$\pm$0.00}& 0.67$\pm$0.03 & 0.78$\pm$0.00 & 0.81$\pm$0.01 \\

1-10 & 2-shot & custom & 0.50$\pm$0.0 & 0.50$\pm$0.00 & 0.50$\pm$0.00 & 0.51$\pm$0.00 & 0.50$\pm$0.00 \\
one-ten & 2-shot & custom & 0.79$\pm$0.00& \textbf{1.00$\pm$0.00}& 0.66$\pm$0.01 & 0.72$\pm$0.00 & 0.79$\pm$0.00 \\

\bottomrule
\end{tabular}

\end{table*}
\begin{table*}[th]
\centering
\caption{Ablation results for the book environment using \textit{Mistral 7B} as our environment. We test the performance of the LLM to give coherent and realistic ratings for user-book interactions. We achieve best overall performance when using 2-shot prompting and our custom system prompt.}
\label{table:ablations_books_prompt_full}
\begin{tabular}{
m{0.05\linewidth}
m{0.07\linewidth}
m{0.065\linewidth}|
m{0.12\textwidth}
m{0.13\textwidth}
m{0.12\textwidth}
m{0.11\textwidth}}
\toprule
\multicolumn{3}{c}{Prompt component} \\
Rating scale & N-shot & System prompt & Category $\uparrow$ & High/low  $\uparrow$ & Collection  \newline of books $\uparrow$ & Agg. score  $\uparrow$ \\
\midrule
1-5 & 0-shot  & default  & 0.68$\pm$0.00 &  \textbf{1.00$\pm$0.00} & 0.65$\pm$0.01 & 0.77$\pm$0.00 \\
1-5 & 0-shot  & custom & 0.83$\pm$0.00 &  \textbf{1.00$\pm$0.00} & 0.68$\pm$0.02 & 0.83$\pm$0.01 \\
1-5 & 1-shot  & default  & 0.87$\pm$0.00 &  \textbf{1.00$\pm$0.00} & 0.81$\pm$0.04 & 0.89$\pm$0.01 \\
1-5 & 1-shot & custom & \textbf{0.89$\pm$0.00} &  \textbf{1.00$\pm$0.00} &  \textbf{0.82$\pm$0.02} &  \textbf{0.90$\pm$0.01} \\
1-5 & 2-shot & default & 0.83$\pm$0.00 & 0.98$\pm$0.00 & 0.73$\pm$0.02 & 0.85$\pm$0.01 \\
1-5 & 2-shot & custom   & 0.85$\pm$0.00 &  \textbf{1.00$\pm$0.00} & 0.76$\pm$0.02 & 0.87$\pm$0.01 \\
\bottomrule
\end{tabular}

\end{table*}

\shortp{High/Low Rating}
We assess whether the environment can accurately infer high ratings for users who provide positive-leaning descriptions, while inferring low ratings for users whose descriptions are negative-leaning. We give each user a set of items and test whether the environment is able to generate high or low ratings, depending on the description of the user.

% We evaluate whether the environment can accurately infer high ratings for users who provide positively inclined descriptions while inferring low ratings for users whose descriptions lean toward negativity. We present a set of items to each user and check whether the model is capable of predicting high and low ratings, respectively. 

%We randomly select a subset of users from our synthetic dataset and fill their history with items from our collection as well as random items
\shortp{Collection of Items} We evaluate the ability of the environment to leverage the historical item ratings of a user to predict their future ratings. We conduct this test by manually selecting a set of item collections belonging to a series (e.g., James Bond, Toy Story, etc.). Subsequently, we randomly select a sample of users from our synthetic dataset and fill their history with items from our collection as well as random items. We assign a high rating to all items in the collection history, and the corresponding average rating to the remaining random items. Success is measured by a high rating for the queried item that is part of the collection. The experiment is repeated by assigning low ratings to the collection items to test the ability of the environment to predict low ratings.

%OLD: We evaluate the ability of the environment to leverage the historical item ratings of a user to predict their future ratings. For this test set, we manually selected a set of item collections that are part of a series and a second set of random items. We select a random subset of users and give them an interaction history. This history contains a permutation of items from the collection together with random items. We set all collection items in the history to a high rating, and the other random items to their average rating. We want to observe a high rating for the queried item that is part of the collection. We then repeat the experiment by setting the collection items to low ratings and check whether model is able to predict a low rating.

\shortp{Similarity to Real Rating Distribution}
% We evaluate if the rating distribution obtained from our environment accurately reflects human behavior by comparing it to the rating distribution of MovieLens and Goodreads \citep{dimitrov2015goodreads}, which are representative samples of human ratings. We sample with replacement from our environment as well as from MovieLens and Goodreads datasets. We then calculate the empirical distribution across these datasets and utilize the total variation distance as a metric to measure similarity.
We evaluate whether the rating distribution obtained from our movie environment accurately reflects human behavior by comparing it to the rating distribution from MovieLens, which are representative samples of human ratings. We sample with replacement from both our environment and the MovieLens dataset. We then compute the empirical distribution across the dataset and use the total variation distance as a metric to measure similarity.  For the book environment, see \ifarxiv\Cref{app:experiment_details}\else Appendix C in \cite{corecco2024suber}\fi.
% we find that existing datasets are too biased towards high ratings to be good candidates.
% \paragraph{Aggregated score}
% The aggregated score is the mean of previous scores, with the minor adjustment that we incorporate one minus the total variation distance in the calculation.
The aggregated score is the mean of all test cases.
All ablations, except where defined otherwise, were performed using the following configurations. We used the 2-shot prompting, a custom system prompt, three item retrievial via T5-similarity, and no reward perturbation. For movies, we used \textit{Mistral 7B} with rating scale 0-9, and for books we use \textit{Mistral 7B} with scale 1-5.

% In the movie environment, we observe the impact of few-shot prompting and our custom prompt on performance (cf.~\Cref{table:ablations_movies_prompt}). The best results emerge from the combination of two-shot prompting and our custom system prompt with a rating scale between 0 and 9, which has the best aggregated score and performance in three out of four test sets. In particular, using a rating scale between 1 and 10 performs poorly, even worse than zero-shot prompting with the default system prompt. Furthermore, prompts using word ratings, while not the best, show improvement compared to the numeric scale from 1 to 10. In \Cref{table:ablations_movies_prompt}, we observe that the best environment configuration achieves 82\% accuracy for the movie collection test set, and 69\% accuracy on the genres score, demonstrating the ability of the environment to capture human concepts such as genres and movie franchises. Similar trends are evident in the book environment (cf.~\Cref{app:ablations_books_full_results}). In general, we observe that larger models perform better across model families (cf.~\Cref{fig:ablations_movies_models,fig:ablations_books_models}). Furthermore, a comparison between \textit{Vicuna-1.5} and \textit{Llama-2-Chat} reveals how fine-tuning the same foundation model (Llama-2) can influence performance.

\begin{figure}[b]
    \includegraphics[width=\linewidth]{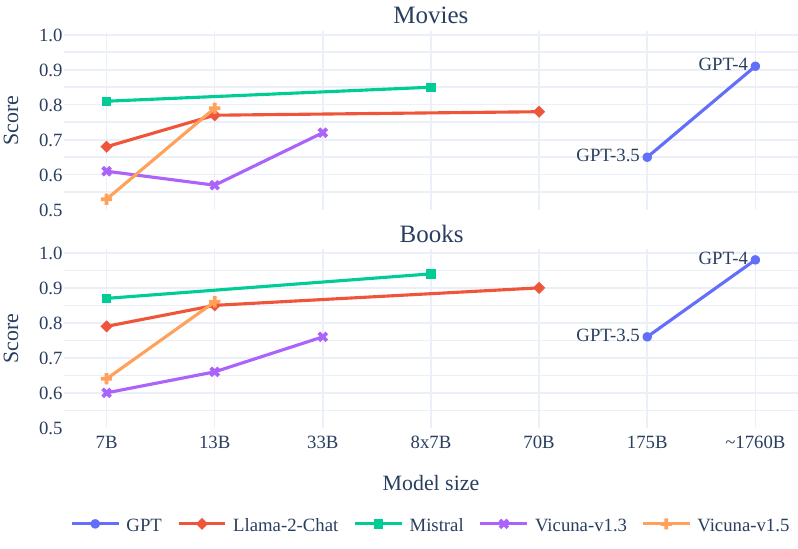}
\caption{Aggregated score across LLM families for the movie environment (top), and for the book environment (bottom) by varying only the LLM component. For details see \ifarxiv\Cref{app:ablations_all_extended}\else Appendix D in \cite{corecco2024suber}\fi.}
       \label{fig:ablations_movies_models}
\end{figure}

% \begin{table}[t]
%     \centering
%   \caption{Comparison of different retrieval strategies on the \textit{collection of items} test set for movies and books. We only show the entries of the test case where the result changed significantly, for extended results see \Cref{app:ablations_all_extended}.}
%     \label{table:ablations_retrieval}
%         \begin{tabular}{     
%          m{0.15\textwidth}
%          m{0.1\textwidth}
%          m{0.1\textwidth}
%          }
%             \toprule
%             {Retrieval \newline component} & Collection of movies  $\uparrow$ & Collection of  books  $\uparrow$ \\
%             \midrule
%             None & 0.49$\pm$0.01 & 0.50$\pm$0.00\\
%             Most recent & 0.62$\pm$0.01 & 0.64$\pm$0.01 \\
%             T5 similarity & 0.67$\pm$0.03 & 0.76$\pm$0.02 \\
%             Feature similarity & \textbf{0.68$\pm$0.01} &  \textbf{0.79$\pm$0.01} \\
%             \bottomrule
%                     \end{tabular}
  
% \end{table}

\shortp{Results} 
In the movie environment, we observe that different prompt strategies generally do not differ significantly from each other in the case of Mistral, with the only two exceptions being the 0-shot prompt with the default system prompt, which performs slightly worse, and the weak performance of the 1-10 rating scale due to tokenization ambiguity. Vicuna, on the other hand, is more affected by different prompt strategies, as shown in \ifarxiv\Cref{app:ablations_movies_full_results}
\else Appendix D in \cite{corecco2024suber}\fi. 
\Cref{table:ablations_movies_prompt} shows the general trend on how the environment can capture human concepts such as genres and movie franchises.
For the book environment (cf.~\Cref{table:ablations_books_prompt_full}) it can be observed that the use of few-shot prompting, as well as the custom system prompt, has a positive impact on the different test cases. Additionally, similar to the movie environment, the model is also able to understand human concepts in the book domain.
In general, we observe that larger models perform better across model families (cf.~\Cref{fig:ablations_movies_models}). In addition, we can see how Mistral performs best among open-source models.

Our ablation of the retrieval component demonstrates that this component plays a crucial role in understanding user interests \ifarxiv(cf.~\Cref{table:ablations_movies_retrival_full,table:ablations_books_retrival_full} in \Cref{app:ablations_all_extended})
\else
(cf. Tables 8 and 12 in the supplementary material \cite{corecco2024suber})
\fi
. Furthermore, the recency approach proves inadequate, while the best-performing retrieval approach is predicated on the similarity of item features. 

% We can also observe that the different perturbations slightly affect the similarity with the actual data distribution, especially in the case of the movie environment (cf.~\Cref{table:ablations_perturbator}).

% Performance Metrics of RL Models Trained on SUBER: Mean Average Precision (MAP@10), Mean Reciprocal Rank (MMR@10), Personalization at Top Ten Recommendations (Pers.@10). 'Liked Genres' indicates the proportion of movies in the top ten recommendations aligned with user-preferred genres (see \Cref{app:rl_model_training} for details).
\begin{table*}[t]
    \centering
  %  Average reward, Coverage of the top ten recommendations (Cov.@10), Personalization of the top ten recommendations (Pers.@10),
    \caption{Performance Metrics of RL Models Trained on SUBER:  Mean Average Precision (MAP@10), Mean Reciprocal Rank (MRR@10), Personalization of the top ten recommendations (Pers.@10). ``Liked genres''  indicates the proportion of movies in the top ten recommendations aligned with user-preferred genres (see \ifarxiv\Cref{app:rl_model_training}\else Appendix F in the supplementary material \cite{corecco2024suber}\fi~for details).}
   \label{table:rl_metrics}
      \begin{tabular}{
    m{0.09\textwidth}
    m{0.09\textwidth}
    m{0.13\textwidth}
    m{0.13\textwidth}
    m{0.12\textwidth}
    m{0.1\textwidth}
    m{0.12\textwidth}
      }
                \toprule
                Algorithm & Average \newline reward &  MAP@10 $\uparrow$ &  MRR@10  $\uparrow$ & Pers.@10 $\uparrow$ & \% Liked \newline genres  $\uparrow$ &  \% Disliked \newline genres  $\downarrow$ \\ 
                \midrule
                DQN & 6.79$\pm$0.06 & 0.53$\pm$0.06 &0.85$\pm$0.06 &0.00$\pm$0.00 &0.42$\pm$0.01 & 0.15$\pm$0.01 \\ 
                PPO &  6.91$\pm$0.03 & 0.59$\pm$0.01 & 0.84$\pm$0.01 & \textbf{0.99$\pm$0.00} & 0.44$\pm$0.01 & 0.15$\pm$0.00  \\
                TRPO & 7.25$\pm$0.08 & 0.67$\pm$0.06 & 0.91$\pm$0.02 &0.35$\pm$0.06 &0.45$\pm$0.02 & 0.14$\pm$0.01  \\
                A2C & \textbf{7.93$\pm$0.07} & \textbf{0.88$\pm$0.01} & \textbf{0.96$\pm$0.01}  & 0.91$\pm$0.03  & \textbf{0.49$\pm$0.02} & \textbf{0.11$\pm$0.01} \\
                \bottomrule
                \end{tabular}
 
\end{table*}

\subsection{Human Evaluation}
\label{subsec:human_evaluation}

\begin{table}[h]
    \centering
        \caption{Human evaluation scores for various LLMs. }
    \label{tab:human_evaluation}
    \begin{tabular}{l|c}
    \toprule
    LLM & Score $\uparrow$ \\
    \midrule
     Random rating & 2.87$\pm$1.51 \\
     Vicuna 13B & 3.22$\pm$1.32 \\  
     Llama-2-Chat 13B & 3.42$\pm$1.22 \\
     Mistral 7B & 3.80$\pm$1.27 \\
     GPT-4 & \textbf{4.47$\pm$0.77} \\
     \bottomrule
    \end{tabular}

\end{table}

We conducted a case study to better evaluate the quality of different LLMs in the rating simulation task. For the study, we sampled ten user-movie interactions, for each interaction we queried four different LLMs. The random rating in \Cref{tab:human_evaluation} serves as a baseline comparison, allowing us to compare the quality of our proposed approach with a random signal. The answer is constructed by sampling a rating uniformly at random between 1 and 10, and having the LLM (\textit{Vicuna 13B}) generate the explanation for the rating.
We then asked participants to rate the quality of the LLM's response on a scale of 1 to 5.

% Participants in this study were recruited from colleagues and consented to be part of this study. The study organizers and participants adhered to strict randomized double-blind procedures. Since this user study does not involve tracking or studying participants our ethics board did not require us to obtain approval.
Participants in this study were recruited from among our colleagues and provided informed consent to participate. The study was designed with strict adherence to randomized double-blind procedures to ensure impartiality and reliability of the results. As this user study did not involve ongoing follow-up or monitoring of the participants, our institutional review board (IRB) determined that formal approval was not required.
From the survey (cf. \Cref{tab:human_evaluation}), we find that users agree more with \textit{GPT-4}, outperforming all other models. Furthermore, we find that \textit{Mistral 7B} is the best LLM among open-source models despite only having 7B parameters. More information on the study setting is provided in \ifarxiv\Cref{app:user_study}\else Appendix E in \cite{corecco2024suber}\fi.

\subsection{Benchmarks}
\label{subsec:experiments_rl}
% \paragraph{RL model}
\label{sec:RL_models}

\begin{figure}[b]
   \centering
   \includegraphics[width=\linewidth]{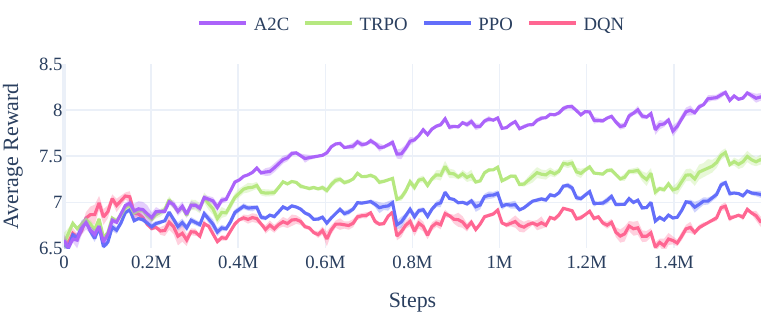}
   \caption{Training plot of various RL models. The y-axis displays the average reward from evaluation samples.}
    \label{fig:reward_during_training}
\end{figure}

We demonstrate the viability of our environment to train an RL recommender system. The architecture of the RL model is inspired by the principles of Low-Rank Approximations in collaborative filtering \citep{aggarwal2016recommender}.
% We demonstrate the viability of our environment for training an RL recommender system. 
We implemented four different agents based on A2C \citep{mnih2016asynchronous}, PPO \citep{schulman2017proximal}, TRPO \citep{schulman2015trust}, and DQN \citep{mnih2013playing}. We train all models for 1.6M steps on SUBER. 
% See \Cref{app:rl_model_training} for more details.
Due to space constraints, a more detailed discussion on the training of reinforcement learning models is deferred to \ifarxiv\Cref{app:rl_model_training}\else Appendix F in the supplementary material \cite{corecco2024suber}\fi.
In addition to using classical RecSys metrics, like $\text{MAP}@10$, $\text{MRR}@10$, and personalization, we introduce two additional metrics to 
% evaluate how well the agent recommends items which align to the preferences of users. 
assess the alignment of the agent's recommendations with user preferences (See \ifarxiv\Cref{app:rl_model_training}\else Appendix F in \cite{corecco2024suber}\fi~for detail metric definitions).
Each user in the training dataset has both preferred and disliked movie genres. 
Based on these data,
the trained RL model generates a list of top-5 movie recommendations for each user:  \textit{percentage liked genres} and \textit{percentage disliked genres }. The recommendations are classified into three categories: \textit{liked} (movies matching preferred genres and excluding disliked ones), \textit{disliked} (movies with disliked genres and without preferred ones), and \textit{neutral} (remaining recommendations).

Our evaluation indicates that the A2C algorithm demonstrates the best overall performance in our case study (cf.~\Cref{table:rl_metrics}). 
Although the PPO algorithm registers a higher personalization score, indicative of its ability to tailor recommendations, it is less effective than A2C in aligning recommendations with user interests, as reflected in the percentage of liked genres metric.  This suggests that A2C is more adept at discerning and catering to user preferences.

\section{Future Work}
One promising direction is to fine-tune the LLM with human feedback to improve the simulated user behavior. This can be achieved using datasets like MovieLens, which provide a natural reward function for RL methods. For instance, the negative squared difference between the LLM rating and the actual rating can be used as a reward.
Currently, the setup considers only static users. Future work could model user evolution over time to reflect changing interests, making synthetic users more realistic and dynamic.
Additionally, exploring ways to enrich the feature space of the LLM could be valuable. By incorporating complex features such as item seasonality and user context, RL models could better capture user behavior, leading to more accurate simulations.

\section{Conclusion}
\label{sec:future_work_and_conclusion}

Our research offers a possible avenue to address the persistent challenge of training recommender systems in the absence of real user interactions. 
Conventional approaches that depend on user-item interaction histories or synthetic data have often failed to replicate real-world usage scenarios accurately.
By introducing SUBER, a novel RL environment designed specifically for recommender system training, and incorporating recent advances in LLMs to emulate human behavior in the training environment, we have proposed a potential solution to this long-standing issue. Our results, as demonstrated through a series of ablation studies, experiments, and human evaluation, underscore the efficacy of our approach. We believe that this work marks a step toward achieving more realistic and practical training environments for recommender systems, even when direct user interactions are unavailable.

%%% Use this command to include your bibliography file.
\clearpage
\bibliography{m1200}

\ifarxiv
\include{appendix}
\else
\fi
\end{document}

%% file: appendix.tex
\appendix

% \section{Prompting rating}
% \subsection{Ratings and tokens}
% \label{app:ratings_and_tokens}
% \comment{Probably we should remove this}
% We decided to adjust the rating scale to range from 0 to 9 instead of 1 to 10. This choice primarily stems from our motivation to address the tokenization ambiguity problem. This problem arises due to the fact that the number 10 can be represented in two distinct ways: by using a token that directly represents the number ten or by using tokens \texttt{1} followed by \texttt{0}. We also experimented using the words  `one' through `ten', this improves slightly on the genres test set, but perform very badly on the similarity score to MovieLens distribution as showed in \Cref{table:ablations_movies_prompt}.

% We also tested using digits from one to ten, and in case the LLM generates the token corresponding to ``1" we sample an additional token to check whether it is ``0". However, we did not continue with this strategy as it yielded poor performance. 

% \subsection{Few-shot prompting example}
% \label{app:few_shot_example}

\clearpage
\section{Details of Movie Environment}
\label{app:movies_env}

\subsection{Users Generation}
When generating synthetic users, our process begins by randomly sampling an age from a distribution reflecting age demographics in the United States~\citep{age_statistic}. In addition, we randomly select a hobby and a profession from predefined lists. These hobby lists are divided into two categories: one tailored to children (aged 4-17) and another for adults (aged 18-75). Users not of working age are assigned the profession ``student," while those of retirement age are categorized as ``retired."

In total, children users can have one of 33 hobbies, while adult users have a choice of 422 hobbies. With regard to professions, there are 200 different options available. Once all user attributes are determined, they are incorporated into a prompt that generates a user description (see \Cref{listing:prompt} for an example). All lists are generated using the \textit{GPT-3.5} model, with the exception of the list of adult hobbies, for which we used data from \citep{Raj_2022}.

For a complete list of hobbies and professions generated, refer to \Cref{app:list}.
An illustrative example of a synthetic user can be found in \Cref{listing:prompt}.

\begin{figure}[b]
\centering
     \includegraphics[width=\linewidth]{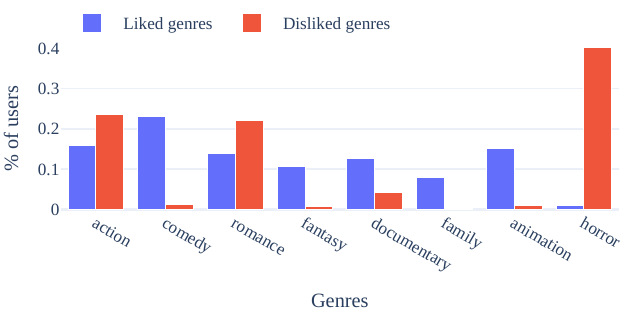}
\caption{Genre preferences of user generated via LLM. For each movie genre, we show in blue the percentage of generated users who like the genre. Similarly, we show in red the percentage of users who do not like the genre.}
  \vspace{2em}
\label{fig:genres_dist_LLM}
\end{figure}

\begin{figure}
\centering
     \includegraphics[width=\linewidth]{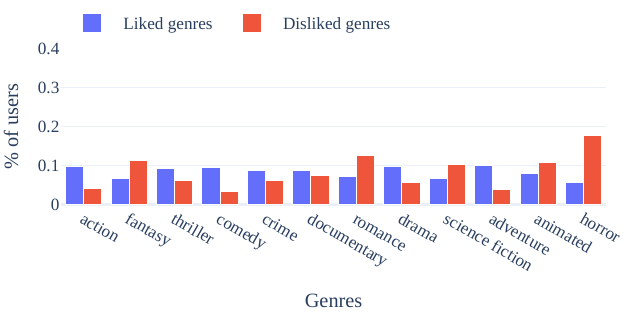}
\caption{Real distribution of genre preferences. For each movie genre, we show in blue the percentage of generated users who like the genre. Similarly, we show in red the percentage of users who do not like the genre.}
\label{fig:genres_dist_sampled}
\end{figure}

To train the RL model, we created an additional dataset using a similar approach. The primary distinction lies in how we sampled the user's preferred and disliked genres, which were not generated using the LLM. This modification was made to ensure that the dataset includes users with a more diverse range of preferences. We sampled the preference for movie genres according to
the distribution of preferred genres in the US in 2018 \citep{consult2018most}.
We show in \Cref{fig:genres_dist_LLM,fig:genres_dist_sampled} how this strategy affects the preference of users for the genre.

\subsection{Items}
\label{app:movie_dataset}

We use the same set of movies as contained in MovieLens  (\textit{ml-latest-small}) \citep{harper2015movielens} for our experiments and collect the respective movie features from TMDB, and show which features we use in \Cref{tab:movie_features}.

\begin{table}[h]
    \centering    
     \caption{For each movie we retrieve the features shown in the table, in our implementation we use only a subset to describe an item.} 
    \label{tab:movie_features}
    \begin{tabular}{rcl}
        \toprule
        Feature & Used & Notes \\
        \midrule
        Actors & Yes &2 principal actors \\
        Budget & No & \\
        Director & Yes & Only for feature similarity\\
        Original language& No & \\
        Original title& No  & \\
        Overview  & Yes & Story-line of the movie\\
        Popularity  & No & \\
        Release date  & Yes & \\
        Revenue  & No &\\
        Runtime  & No &\\
        Title  & Yes &\\
        TMDB ID & No & Unique id in the dataset\\
        Vote average  &  Yes & \\
        Vote count  & No & \\
        \bottomrule
    \end{tabular}
   
\end{table}

\subsection{Prompting}
In this section, we provide several examples of different prompting strategies. We primarily focus on three key approaches: using prompts with digits ranging from 1 to 10, utilizing prompts with digits from 0 to 9, and employing prompts with word representations for numbers from one to ten.

Both the approach of using digits from 0 to 9 and the word-based approach are designed to address tokenization ambiguity. This ambiguity arises because the number 10 can be tokenized in two different ways: as the token ``10" directly or as separate tokens ``1" and` `0." We also explored generating numbers directly without restricting them to a single token in the 1-10 approach. However, this approach exhibited poor performance, which led us to refrain from further experimentation.

\begin{lstlisting}[style=interaction, label=listing:example_interaction_0_9, caption={Example query for rating to the LLM (\textit{Vicuna-v1.5-13B}) using the 0-9 scale.}]
[system prompt]
[few shot prompts]
Q: <input>Emily</input> is a <input>37</input> years old <input>woman</input>, she is <input>a detective, she has a hobby of collecting compact discs, she likes to watch romance and horror movies in her free time, she dislikes action and comedy movies because she finds them too chaotic and not interesting, her secondary hobbies are reading mystery novels and playing the piano.</input>
<input>Emily</input> has previously watched the following movies (in parentheses are the ratings she gave on a scale of 0 to 9): "<input>Twelve Monkeys</input>" (<input>7</input>), "<input>Star Wars</input>" (<input>6</input>), "<input>Top Gun</input>" (<input>5</input>).
Consider the movie "<input>Broken English</input>", released in <input>1996</input>, which is described as follows: <input>Ivan is the fierce patriarch of a family of Croatian refugees living in Auckland during the Yugoslav wars. Nina is his daughter, ready to live on her own, despite his angry objections. Eddie is the Maori she takes as her lover. Nina works at the restaurant where Eddie cooks. For a price, she agrees to marry another restaurant employee, a Chinese man, so that he can establish permanent residency. The money gives her the independence she needs to leave her parents' house and move in with Eddie. Complications arise when Eddie realizes the depth of her father's fury and the strength of Nina's family ties.</input> The movie "<input>Broken English</input>" contains the following genres:<input>
-romance
-drama</input>
Here are the 2 main actors of the movie, in order of importance: <input>Rade SerbedZija (M), Aleksandra Vujcic (F)</input>. On average, people rate the movie "<input>Broken English</input>" <input>5.6</input> on a scale of 0 to 9. <input>Emily</input> watches the movie "<input>Broken English</input>" for the <input>1st</input> time.
What can you conclude about <input>Emily</input>'s rating for the movie "<input>Broken English</input>" on a scale of 0 to 9, where 0 represents a low rating and 9 represents a high rating, based on available information and logical reasoning? 

A: Based on <input>Emily</input>'s preferences and tastes, I conclude that she will assign a rating of <llm>7</llm>
\end{lstlisting}

\begin{lstlisting}[style=interaction, label=listing:example_interaction_one_ten, caption={Example query for rating to the LLM (\textit{Vicuna-v1.5-13B}) using rating scale one-ten.}]
[system prompt]
[few shot prompts]
Q: <input>Emily</input> is a <input>37</input> years old <input>woman</input>, she is <input>a detective, she has a hobby of collecting compact discs, she likes to watch romance and horror movies in her free time, she dislikes action and comedy movies because she finds them too chaotic and not interesting, her secondary hobbies are reading mystery novels and playing the piano.</input>
<input>Emily</input> has previously watched the following movies (in parentheses are the ratings she gave on a scale of 1 to 10): "<input>Twelve Monkeys</input>" (<input>8</input>), "<input>Star Wars</input>" (<input>7</input>), "<input>Top Gun</input>" (<input>6</input>).
Consider the movie "<input>Broken English</input>", released in <input>1996</input>, which is described as follows: <input>Ivan is the fierce patriarch of a family of Croatian refugees living in Auckland during the Yugoslav wars. Nina is his daughter, ready to live on her own, despite his angry objections. Eddie is the Maori she takes as her lover. Nina works at the restaurant where Eddie cooks. For a price, she agrees to marry another restaurant employee, a Chinese man, so that he can establish permanent residency. The money gives her the independence she needs to leave her parents' house and move in with Eddie. Complications arise when Eddie realizes the depth of her father's fury and the strength of Nina's family ties.</input> The movie "<input>Broken English</input>" contains the following genres:<input>
-romance
-drama</input>
Here are the 2 main actors of the movie, in order of importance: <input>Rade SerbedZija (M), Aleksandra Vujcic (F)</input>. On average, people rate the movie "<input>Broken English</input>" <input>6.6</input> on a scale of one to ten. <input>Emily</input> watches the movie "<input>Broken English</input>" for the <input>1st</input> time.
What can you conclude about <input>Emily</input>'s rating for the movie "<input>Broken English</input>" on a scale of one to ten, where one represents a low rating and ten represents a high rating, based on available information and logical reasoning? 

A: Based on <input>Emily</input>'s preferences and tastes, I conclude that she will assign a rating of <llm>eight</llm> 
\end{lstlisting}

\subsubsection{Custom System Prompt}
\label{subsubsec:system_prompt_movie}

We also experimented with various system prompts, which are predefined text or instructions used to initiate a conversation or request from a user when interacting with a language model. The primary objective is to encourage the model to generate ratings that are less biased and more closely aligned with the information provided to the model. This includes factors such as the user description, the list of movies watched previously, and the overview of the queried movie, all of which play a role in shaping the predictions of a model. In \Cref{listing:our_system_prompt_app}, we present our customized system prompt utilized for various analyses in \Cref{sec:experiments}.

\begin{lstlisting}[style=interaction, label=listing:our_system_prompt_app, caption={An advanced system prompt guiding the model to provide personalized and unbiased movie ratings based on detailed user and movie data.}]
<template>You are a highly sophisticated movie rating assistant, equipped with an advanced understanding of human behavior. Your mission is to deliver personalized movie recommendations by carefully considering the unique characteristics, tastes, and past-seen films of each individual. When presented with information about a specific movie, you will diligently analyze its plot, primary genres, actors, and average rating. Using this comprehensive understanding, your role is to provide thoughtful and accurate ratings for movies on a scale of 1 to 10, ensuring they resonate with the person's preferences and cinematic inclinations. Remain impartial and refrain from introducing any biases in your predictions. You are an impartial and reliable source of movie rating predictions for the given individual and film descriptions.</template>
\end{lstlisting}

\subsubsection{Query Template}
\label{app:query_template_movie}
In the following section, we provide an example prompt and accompanying LLM answer. It is important to note that \Cref{listing:example_interaction_1} displays the complete response from the model, not just the rating. During interaction with an RL model, we halt generation after producing the rating.
\begin{lstlisting}[style=interaction, label=listing:example_interaction_1, caption={Example query for rating to the LLM (\textit{Vicuna-v1.5-13B}). For each user we inject their description, which contains preferences and tastes. Then we provide the movie details: storyline, genres, main actors and vote average.}]
[system prompt]
[few shot prompts]
Q: <input>Emily</input> is a <input>37</input> years old <input>woman</input>, she is <input>a detective, she has a hobby of collecting compact discs, she likes to watch romance and horror movies in her free time, she dislikes action and comedy movies because she finds them too chaotic and not interesting, her secondary hobbies are reading mystery novels and playing the piano.</input>
<input>Emily</input> has previously watched the following movies (in parentheses are the ratings she gave on a scale of 1 to 10): "<input>Twelve Monkeys</input>" (<input>8</input>), "<input>Star Wars</input>" (<input>7</input>), "<input>Top Gun</input>" (<input>6</input>).
Consider the movie "<input>Broken English</input>", released in <input>1996</input>, which is described as follows: <input>Ivan is the fierce patriarch of a family of Croatian refugees living in Auckland during the Yugoslav wars. Nina is his daughter, ready to live on her own, despite his angry objections. Eddie is the Maori she takes as her lover. Nina works at the restaurant where Eddie cooks. For a price, she agrees to marry another restaurant employee, a Chinese man, so that he can establish permanent residency. The money gives her the independence she needs to leave her parents' house and move in with Eddie. Complications arise when Eddie realizes the depth of her father's fury and the strength of Nina's family ties.</input> The movie "<input>Broken English</input>" contains the following genres:<input>
-romance
-drama</input>
Here are the 2 main actors of the movie, in order of importance: <input>Rade SerbedZija (M), Aleksandra Vujcic (F)</input>. On average, people rate the movie "<input>Broken English</input>" <input>6.6</input> on a scale of 1 to 10. <input>Emily</input> watches the movie "<input>Broken English</input>" for the <input>1st</input> time.
What can you conclude about <input>Emily</input>'s rating for the movie "<input>Broken English</input>" on a scale of 1 to 10, where 1 represents a low rating and 10 represents a high rating, based on available information and logical reasoning? 

A: Based on <input>Emily</input>'s preferences and tastes, I conclude that she will assign a rating of <llm>8</llm> on a scale of 1 to 10, because <llm>
1) she has previously rated movies within the same genre, such as "Twelve Monkeys," which also falls under drama and romance categories, and gave it a higher score (8 out of 10); 
2) Her secondary hobbies include reading mystery novels and playing the piano, suggesting an appreciation for more introspective and emotionally engaging content; 
3) The plot of "Broken English" revolves around family dynamics, cultural clashes, and personal growth, all of which align with her interests in watching romance and drama films. Although Emily doesn't typically enjoy action or comedy movies, this does not necessarily mean that she would dislike "Broken English." Instead, it suggests that she might find other genres less appealing due to their lack of depth. Overall, considering her previous ratings and general preferences, it is reasonable to assume that Emily would rate "Broken English" highly.</llm>
\end{lstlisting}

\subsubsection{N-Shot Prompting}
In this section, we present two illustrative examples (\Cref{listing:1_shot_prompt_movies} and \ref{listing:2_shot_prompt_movies}) used for both one-shot and two-shot prompts. These examples adhere to the query template shown in \Cref{app:query_template_movie}, guiding the model on how to formulate responses considering all the features. In addition, both examples provide comprehensive and detailed answers.

One advantage of few-shot prompting is that the examples are given at the start of the prompt. This gives us the opportunity to use the key-value cache \citep{pope2023efficiently} without slowing down the environment.

Furthermore, these two examples are designed to cover both positive and negative rewards. In the first example, the user rates the queried movie highly, while in the second example, the user rates the queried movie poorly, thereby encompassing a range of user preferences and responses.

\begin{lstlisting}[style=interaction, label=listing:1_shot_prompt_movies, caption={First shot example for movies environment.}]
Q: Alex is a 12 years old boy, he is captivated by space exploration. With wide eyes and endless wonder, he devours books on galaxies and dreams of becoming an astronaut. Nights find him gazing at the stars, his imagination soaring with each cosmic discovery. Alex's passion for space knows no bounds as he reaches for the stars.
Alex has previously watched the following movies (in parentheses are the ratings he gave on a scale of 1 to 10): "Madagascar" (7), "Foodfight!" (2), "Interstellar" (10).
Consider the movie "Zootopia", released in 2016, which is described as follows: determined to prove herself, Officer Judy Hopps, the first bunny on Zootopia's police force, jumps at the chance to crack her first case - even if it means partnering with scam-artist fox Nick Wilde to solve the mystery. The movie "Zootopia" contains the following genres:
-animation
-adventure
-family
-comedy
Here are the 2 main actors of the movie, in order of importance: Jason Bateman (M), Ginnifer Goodwin (F). On average, people rate the movie "Zootopia" 7.7 on a scale of 1 to 10. Alex watches the movie "Zootopia" for the 1st time.
What can you conclude about Alex's rating for the movie "Zootopia" on a scale of 1 to 10, where 1 represents a low rating and 10 represents a high rating, based on available information and logical reasoning? 
A: Based on Alex's preferences and tastes, I conclude that he will assign a rating of 9 on a scale of 1 to 10, because, even though the movie is not space-related, he previously enjoyed watching "Madagascar" and gave it a high rating. Since "Zootropolis" shares many similarities with "Madagascar," such as being animated movies with animals living in a society resembling humans, it is highly likely that Alex also likes "Zootropolis." Both films explore themes of friendship, identity, and the challenges of coexistence within diverse communities. Although "Zootropolis" doesn't have a direct connection to space, Alex still appreciates it and gives it a high rating. Furthermore, considering that children generally have a fondness for animated movies, it is reasonable to assume that Alex would rate "Zootropolis" high." Additionally, "Zootropolis" is generally favored by a larger audience compared to "Madagascar," further supporting the idea that Alex would give it a higher rating.
\end{lstlisting}

\begin{lstlisting}[style=interaction, label=listing:2_shot_prompt_movies, caption={Second shot example for movies environment.}]
Q: Nicholas is a 26 years old man, he is a thrill-seeker who loves action and adventure and doesn't care much for romance. He enjoys activities like rock climbing, extreme sports, and exciting trips that make his heart race. As a wilderness guide, he leads others through tough terrains and shows them the beauty of wild places. During his free time, Nicholas loves reading gripping books that take him on exciting journeys. He looks for friends who also enjoy thrilling experiences and share his passion for living life to the fullest. Adrenaline gives him an amazing feeling, and he prefers action and adventure over romance any day.
Nicholas has previously watched the following movies (in parentheses are the ratings he gave on a scale of 1 to 10): "Fifty Shades of Grey" (3), "John Wick" (9), "The Dark Knight" (10).
Consider the movie "La La Land", released in 2016, which is described as follows: mia, an aspiring actress, serves lattes to movie stars in between auditions and Sebastian, a jazz musician, scrapes by playing cocktail party gigs in dingy bars, but as success mounts they are faced with decisions that begin to fray the fragile fabric of their love affair, and the dreams they worked so hard to maintain in each other threaten to rip them apart. The movie "La La Land" contains the following genres:
-comedy
-drama
-romance
-music
Here are the 2 main actors of the movie, in order of importance: Ryan Gosling (M), Emma Stone (F). On average, people rate the movie "La La Land" 7.9 on a scale of 1 to 10. Nicholas watches the movie "La La Land" for the 1st time.
What can you conclude about Nicholas's rating for the movie "La La Land" on a scale of 1 to 10, where 1 represents a low rating and 10 represents a high rating, based on available information and logical reasoning? 
A: Based on Nicholas's preferences and tastes, I conclude that he will assign a rating of 4 on a scale of 1 to 10, because Nicholas has a strong preference for adrenaline-inducing action, thriller, and horror movies, he would likely rate the movie "La La Land" 4 out of 10. This is evident from the description of Nicholas, which highlights his enthusiasm for action-packed films that provide a surge of thrill and excitement. In his previous film ratings, action movies generally received higher scores, while films that didn't offer the same adrenaline rush, like "Fifty Shades of Grey", received lower ratings, such as a 3. As "La La Land" is a romantic musical and not focused on action, it may not resonate as strongly with Nicholas's taste for thrilling experiences. While the film is generally well-liked with an average rating of 7.9, Nicholas's preference for adrenaline-filled plots might lead him to rate "La La Land" lower than the overall community rating. However, it's likely that he wouldn't rate it as low as "Fifty Shades of Grey" due to its higher popularity and appreciation among viewers who enjoy romance and musical genres.
\end{lstlisting}

%%%%%%%%%%%%%%%%%%%%%%%%%%%%%%%%%%%%%%%%%%%%%%%%%%%%%%%%%%%%%%%%%%%%%%%%%%%%%%%%%%%%%%%%%%%%%%%%%%
\section{Details Books Environment}
\label{app:books_env}

\subsection{Users Generation}
We generate the user dataset in the same way as we did for the users in the movie dataset, by sampling the user features from the same lists using the same method.

\subsection{Items}
We filter books from the Amazon Book Dataset \citep{ni-etal-2019-justifying} by removing books that did not have all the features: categories, description, title, and publication date.
We also limit the categories to those with at least 100 books, so we do not get fine-grained categories.

\subsection{Prompting}
\subsubsection{Custom System Prompt}
\label{subsubsec:system_prompt_books}
We also experimented with various system prompts, which are predefined text or instructions used to initiate a conversation or request from a user when interacting with a language model. The primary objective is to encourage the model to generate ratings that are less biased and more closely aligned with the information provided to the model. This includes factors such as the user description, the list of books read previously, and the back-cover of the queried book, all of which play a role in shaping the predictions of a model.

\begin{lstlisting}[style=interaction, label=listing:our_system_prompt_books, caption={An advanced system prompt guiding the model to provide personalized and unbiased movie ratings based on detailed user and movie data.}]
<template>You are a highly sophisticated book rating assistant, equipped with an advanced understanding of human behavior. Your mission is to deliver personalized book recommendations by carefully considering the unique characteristics, tastes, and past read books of each individual. When presented with information about a specific book, you will diligently analyze its backcover, primary category, authors, and average rating. Using this comprehensive understanding, your role is to provide thoughtful and accurate ratings for books on a scale of 1 to 5, ensuring they resonate with the person's preferences and reading inclinations. Remain impartial and refrain from introducing any biases in your predictions. You are an impartial and reliable source of book rating predictions for the given individual and book descriptions.</template>
\end{lstlisting}

\subsubsection{Query Template}
\label{app:query_template_books}
\begin{lstlisting}[style=interaction, label=listing:example_interaction, caption={Example query for rating to the LLM. For each user we inject their description, which contains preferences and tastes. Then we provide the movie details: backcover, category, authors and vote average.}]
[system prompt]
[few shot prompts]
Samuel is a 17 years old boy, he is an apprentice and loves to work with his hands. He is very interested in animal fancy and loves to breed and show his animals. Samuel is very fitness-conscious and loves to stay active. He enjoys hiking and playing sports. Samuel is a big fan of the Spirit category and enjoys reading books that can help him improve his spiritual life. He also loves reading books about crafts and enjoys learning new techniques. Samuel is also very close to his family and enjoys reading books about family relationships. He is not a big fan of religion and finds it to be boring. He also dislikes music, literary collections and juvenile fiction. He finds them to be too slow paced and not interesting enough for him.
Samuel has previously read the following books (in parentheses are the ratings he gave on a scale of 1 to 5): "The Two Towers" (5), "The Fellowship of the Ring" (5), "The Horse and His Boy" (3).
Consider the book "The Return of the King", released in 1955, which is described as follows: one Ring to rule them all, One Ring to find them, One Ring to bring them all and in the darkness bind them. The Dark Lord has risen, and as he unleashes hordes of Orcs to conquer all Middle-earth, Frodo and Sam struggle deep into his realm in Mordor. To defeat Sauron, the One Ring must be destroyed in the fires of Mount Doom. But the way is impossibly hard, and Frodo is weakening. The Ring corrupts all who bear it and Frodo's time is running out.Will Sam and Frodo succeed, or will the Dark Lord rule Middle-earth once more? The book "The Return of the King" belongs to the following categories:
-Fantasy
-Classic
-Fiction
-Adventure
The author of the book is J.R.R. Tolkien. On average, people rate the book "The Return of the King" 4.6 on a scale of 1 to 5. Samuel reads the book "The Return of the King" for the 1st time.
What can you conclude about Samuel's rating for the book "The Return of the King" on a scale of 1 to 5, where 1 represents a low rating and 5 represents a high rating, based on available information and logical reasoning? 
Q: Based on Samuel's preferences and tastes, I conclude that he will assign a rating of <llm>5</llm>
\end{lstlisting}

\subsubsection{N-Shot Prompting}
In this section, we present two illustrative examples (\Cref{listing:1_shot_prompt_books} and \ref{listing:2_shot_prompt_books}) used for both one-shot and two-shot prompts. These examples adhere to the query template shown in \Cref{app:query_template_books}, guiding the model on how to formulate responses by considering all features. Additionally, both examples provide comprehensive and detailed answers.

Furthermore, these two examples are designed to cover both positive and negative rewards. In the first example, the user rates the queried book highly, while in the second example, the user rates the queried book poorly, thereby encompassing a range of user preferences and responses.

\begin{lstlisting}[style=interaction, label=listing:1_shot_prompt_books, caption={First shot example for books environment}]
Q: Emilia is a 20 years old woman, she is an avid reader, she spends much of her free time lost in the pages of books, especially those filled with magical worlds, exciting adventures and tales of elves. Her passion for the magical realms of literature is evident in her vivid imagination and the way her eyes light up when discussing stories. As well as reading, she enjoys drawing, attending book club meetings, stargazing, sipping tea on rainy days, baking and getting lost in stories about elves.
Emilia has previously read the following books (in parentheses are the ratings she gave on a scale of 1 to 5): "Harry Potter and the Chamber of Secrets" (5), "Harry Potter and the Philosopher's Stone" (5), "Eragon" (5).
Consider the book "Harry Potter and the Prisoner of Azkaban", released in 1999, which is described as follows: harry Potter, along with his best friends, Ron and Hermione, is about to start his third year at Hogwarts School of Witchcraft and Wizardry. Harry can't wait to get back to school after the summer holidays. (Who wouldn't if they lived with the horrible Dursleys?) But when Harry gets to Hogwarts, the atmosphere is tense. There's an escaped mass murderer on the loose, and the sinister prison guards of Azkaban have been called in to guard the school... The book "Harry Potter and the Prisoner of Azkaban" belongs to the following categories:
-Fiction
-Young Adult
-Magic
-Classic
The author of the book is J.K. Rowling. On average, people rate the book "Harry Potter and the Prisoner of Azkaban" 4.6 on a scale of 1 to 5. Emilia reads the book "Harry Potter and the Prisoner of Azkaban" for the 1st time.
What can you conclude about Emilia's rating for the book "Harry Potter and the Prisoner of Azkaban" on a scale of 1 to 5, where 1 represents a low rating and 5 represents a high rating, based on available information and logical reasoning? 
A: Based on Emilia's preferences and tastes, I conclude that she will assign a rating of 5 on a scale of 1 to 5, because from Emilia's description we can clearly see her love for magic and fantasy books, moreover the book "Harry Potter and the Prisoner of Azkaban" is the third book of the Harry Potter series, and from her history we can see that she has already read the first two books of the series and she loved them, because she assigend a perfect score of 5. Moreover, the third book that she has read has a lot to do with magic, which underlines her interest in magical words and stories. The book also has a very high average rating, suggesting that people love the book.
\end{lstlisting}

\begin{lstlisting}[style=interaction, label=listing:2_shot_prompt_books, caption={Second shot example for books environment}]
Q: Mary is a 12 years old girl, she is a person with an overflowing heart, shares an extraordinary bond with the animal kingdom. Her eyes light up with wonder at the sight of a furry friend, and her days are filled with joyful adventures exploring the world's wildlife. From rescuing lost kittens to befriending birds in her backyard, Mary's compassion knows no bounds. Her room is a sanctuary of stuffed animals and nature books, a testament to her unwavering love for all creatures great and small. She is afraid of shadows and loves to sleep with the light on.
Mary has previously read the following books (in parentheses are the ratings she gave on a scale of 1 to 5): "Charlotte's Web" (5), "The Shining" (1), "The Trouble with Tuck" (4).
Consider the book "Coraline", released in 2002, which is described as follows: the day after they moved in, Coraline went exploring.... In Coraline's family's new flat are twenty-one windows and fourteen doors. Thirteen of the doors open and close. The fourteenth is locked, and on the other side is only a brick wall, until the day Coraline unlocks the door to find a passage to another flat in another house just like her own. Only it's different. At first, things seem marvelous in the other flat. The food is better. The toy box is filled with wind-up angels that flutter around the bedroom, books whose pictures writhe and crawl and shimmer, little dinosaur skulls that chatter their teeth. But there's another mother, and another father, and they want Coraline to stay with them and be their little girl. They want to change her and never let her go. Other children are trapped there as well, lost souls behind the mirrors. Coraline is their only hope of rescue. She will have to fight with all her wits and all the tools she can find if she is to save the lost children, her ordinary life, and herself. Critically acclaimed and award-winning author Neil Gaiman will delight readers with his first novel for all ages. The book "Coraline" belongs to the following categories:
-Horror
-Fantasy
-Fiction
-Young Adult
The author of the book is Neil Gaiman. On average, people rate the book "Coraline" 4.1 on a scale of 1 to 5. Mary reads the book "Coraline" for the 1st time.
What can you conclude about Mary's rating for the book "Coraline" on a scale of 1 to 5, where 1 represents a low rating and 5 represents a high rating, based on available information and logical reasoning? 
A: Based on Mary's preferences and tastes, I conclude that she will assign a rating of 2 on a scale of 1 to 5 because, although it is a book for children, as it also falls into the Young Adult category, it is not a book that suits Mary's personality well; in fact, she is afraid of shadows when she needs to sleep, which suggests that the book "Caroline", which is mainly a horror book, is not well suited to Mary. Also, given her sensitivity and love of animals, the creepy and potentially frightening aspects of the story are too much for her. We can also see from Mary's previous red books that she has had a bad experience with horror books, in fact she rated "The Shining" 1 out of 5, whereas "Caroline" is more suitable for children, which explains why Mary probably rated "Caroline" 2 while she rated "The Shining" 1. 
\end{lstlisting}

%%%%%%%%%%%%%%%%%%%%%%%%%%%%%%%%%

\section{Experiment Details}
\label{app:experiment_details}
In this appendix we present more details regarding the test cases, showing also the specifics for both implementations of SUBER for movies and books.

\shortp{Genres/Categories}
For the genre test set, we manually created four distinct users for each genre: action, animation, comedy, documentary, family, fantasy, horror, and romance. These users included two women and two men, with one younger individual and one older individual for each gender. In constructing the user descriptions, we ensured that each person consistently rated a specific genre highly (between 8 and 10) while assigning lower ratings to all other genres (between 1 and 5). The asymmetry in high and low ratings is motivated by research done by \citet{ramos2015statistical}, where they analyze rating behavior on IMDB.
We then presented these users with a set of 20 movies of their preferred genres and another 20 movies from genres they dislike. Our evaluation metric is the percentage of successful predictions in these scenarios.

For the book environment, users are created in the same manner with the exception of their genre preferences, which are specific to book categories rather than movies. The available book categories include fiction, biography, economics, health, philosophy, computer, humor, and drama.

\begin{lstlisting}[style=interaction, label=listing:user_genre_description_test, caption={Example description of a user of the genres test set}]
Oliver is a 27 years old man, he is a gentle and introspective man, holds a deep affection for animation films. He possesses a keen eye for detail and an appreciation for the craftsmanship that goes into creating animated works. Oliver's love for animation is evident in his collection of concept art and his fascination with the behind-the-scenes process. Oliver gives only a high rating to animation films, the motivation ies in their ability to convey profound messages in a visually captivating manner. He believes that animation has a unique power to touch the hearts of both children and adults alike. On the other hand Oliver thinks that a film which is not an animated films is not woth watching, since realismus is bad for people, for this reason he assigns a low rating (between 1-5) to every film, which is not an animation film.
\end{lstlisting}

\shortp{High/Low}
We created eight hand-made users: four females and four males, with two young and two elderly individuals of each gender. Within each age group, there is one user who consistently rates items highly and one who consistently rates items low. In this evaluation, we present 160, 20 when using a model based on a paid API (GPT-3.5, GPT-4), items to each of these users, and assess the environment performance by measuring the percentage of successful predictions. The correctness of the environment is determined by its ability to predict high ratings for users whose descriptions explicitly indicate a preference for higher ratings and low ratings for users whose descriptions imply a preference for lower ratings.

\begin{lstlisting}[style=interaction, label=listing:user_high_description_test, caption={Example description of a user of the high/low test set}]
Ava is a 80 years old woman, she is an elderly woman finds great pleasure in reading books, as they are her sole source of passion and entertainment. With no other hobbies to occupy her time, she devotes herself entirely to the world of books. As a token of her appreciation for the writers, she consistently awards a perfect rating of 5 to express her gratitude.
\end{lstlisting}

\shortp{Collection of Items}
To evaluate the movie environment, we took a selection of 22 movie franchises as our test cases. For each franchise, we sample a set of 100 users from our dataset, 50 when using a model based on a paid API (GPT-3.5, GPT-4). To construct the histories of the users, we included all movies from the respective franchise, except one, and filled the histories with additional randomly chosen movies. For every user, we designed two distinct queries for the environment. In the first query, all the movies were ones that the user had rated highly in the past. In contrast, in the second query, the user had assigned low ratings to all the movies. Subsequently, we requested a rating for the movie that had been excluded. For the first type of query, we consider the environment to be successful if the user assigns a high rating (consistent with their previous high ratings). In the second type of query, success is determined by the user assigning a low rating (in agreement with their previous low ratings).

This methodology resulted in the creation of 200 queries for the environment for each franchise. The overall score is calculated based on the percentage of successful predictions in all tests in the 22 different film franchises.
% For more information on the franchises used, see \comment{add appendix}.

We tested the book environment in a similar way, with the only exception that we used 20 book collections.
% \comment{Add table with books collection}

\shortp{Similarity to Real Rating Distribution}
To calculate the similarity with the true data distribution of MovieLens, we begin by sampling two datasets, $D_E$ (for our environment) and $D_M$ (for MovieLens), with replacement. Our sampling process is as follows: for the movie environment, we randomly select a user and a movie and request the rating. Similarly, for MovieLens, we start by choosing a movie uniformly at random  and then choose one of its ratings randomly.

Let $D_E = \{(m_1, u_1, r_1), \dots,(m_{N}, u_{N}, r_{N}) \}$ be the dataset sampled from our environment, and let $D_M = \{(m_1', u_1', r_1'), \dots,(m_{N'}', u_{N'}', r_{N'}') \}$ be the MovieLens dataset. Where a triplet $(m,u,r)$ represents a user's rating of movie $m$ with a score of $r$.

We compute empirical rating distributions for both MovieLens and the movie dataset from our environment as follows:
\begin{align*}
&p_{D_M}(j) = \frac{\vert \{(m,u,r) \in D_M \mid r = j \}\vert}{\vert D_M\vert }\\
&p_{D_E}(j) = \frac{\vert \{(m,u,r) \in D_E \mid r = j \}\vert}{\vert D_E\vert }
\end{align*}

To compare these two distributions, we calculate the total variation distance between the discrete probability distributions $p_{D_M}$ and $p_{D_E}$:
\begin{equation}
\delta(p_{D_M},p_{D_E}) = \frac{1}{2}\|p_{D_M}-p_{D_E}\|_1 = \frac{1}{2}\sum_{j \in [10]}|p_{D_M}(j) - p_{D_E}(j)|.
\end{equation}
We then compute the similarity using the variation distance as follows:
\begin{equation}
    \text{sim}(D_M, D_E) = 1 - \delta(p_{D_M}, p_{D_E}).
\end{equation}
% For the book environment we proceed similarly with the only difference that we used Goodreads as a comparison.

For the book environment, we find that existing datasets are too biased toward high ratings to be good candidates, so the aggregated score does not include this test case.

\section{Extended Ablations Results}
\label{app:ablations_all_extended}

\label{app:ablations_movies_full_results}
In the following section, we present additional results for both the movie and book environments. We show more detailed scores for the different test set of the ablation for the different models, as well as the impact of the different perturbator and retrieval component on the scores of \textit{Mistral 7B}. Additionally we provide more detail on the impact of different prompting strategies for \textit{Vicuna-v1.5-13B} in the movie setting
\begin{table*}[h]

\caption{Ablation results for the movie environment using the following settings: 0-9 rating scale, 2-shot, custom system prompt, \textit{T5-similarity} movie retrieval, and perturbation \textit{none}. For each prompt component, we show the aggregated score and specific subscores for the various test cases.}
\label{table:ablations_movies_models_full}
\begin{tabular}{m{0.15\textwidth}m{0.07\textwidth}
m{0.12\textwidth}
m{0.13\textwidth}
m{0.12\textwidth}
m{0.10\textwidth}
m{0.11\textwidth}}
\toprule
LLM & Size & Genres $\uparrow$ & High/Low $\uparrow$  & Collection \newline of movies $\uparrow$ & Similarity \newline to ML $\uparrow$ & Agg. score $\uparrow$  \\
\midrule
  GPT-4 & 1760B & \textbf{0.96$\pm$0.00} & \textbf{1.00$\pm$0.00} & \textbf{0.98$\pm$0.00} & 0.69$\pm$0.04 & \textbf{0.91$\pm$0.01} \\
  GPT-3.5 & 175B & 0.66$\pm$0.00 & 0.94$\pm$0.00 & 0.50$\pm$0.00 & 0.49$\pm$0.02 & 0.65$\pm$0.00 \\
 \midrule
 \multirow{3}{*}{Llama-2-Chat} & 70B & 0.80$\pm$0.00 & 1.00$\pm$0.00 & 0.67$\pm$0.01 & 0.66$\pm$0.00 & 0.78$\pm$0.00 \\
 & 13B   & 0.76$\pm$0.00 &  \textbf{1.00$\pm$0.00} & 0.72$\pm$0.01 & 0.60$\pm$0.00 & 0.77$\pm$0.00 \\
    & 7B & 0.55$\pm$0.01 &  \textbf{1.00$\pm$0.00} & 0.72$\pm$0.02 & 0.47$\pm$0.00 & 0.68$\pm$0.00 \\
 \midrule
 \multirow{3}{*}{Vicuna-v1.3} & 33B & 0.48$\pm$0.00 &  \textbf{1.00$\pm$0.00} & 0.78$\pm$0.02 & 0.64$\pm$0.00 & 0.72$\pm$0.01 \\
  & 13B & 0.42$\pm$0.01 & 0.62$\pm$0.02 & 0.64$\pm$0.01 & 0.59$\pm$0.00 & 0.57$\pm$0.01 \\
   & 7B & 0.37$\pm$0.00 & 0.85$\pm$0.01 & 0.61$\pm$0.02 & 0.61$\pm$0.00 & 0.61$\pm$0.01 \\
 \midrule
 \multirow{2}{*}{Vicuna-v1.5} & 13B & 0.69$\pm$0.00 &  \textbf{1.00$\pm$0.00} & 0.77$\pm$0.02 & 0.68$\pm$0.00 & 0.79$\pm$0.01 \\

 & 7B & 0.44$\pm$0.00 & 0.34$\pm$0.00 & 0.73$\pm$0.04 & 0.63$\pm$0.00 & 0.53$\pm$0.01 \\
  \midrule
 \multirow{2}{*}{Mistral}  & 8x7B & 0.81$\pm$0.01 & 1.00$\pm$0.00 & 0.78$\pm$0.03 & \textbf{0.82$\pm$0.00} & 0.85$\pm$0.01 \\
 & 7B & 0.79$\pm$0.00 &  \textbf{1.00$\pm$0.00} & 0.67$\pm$0.03 & 0.78$\pm$0.00 & 0.81$\pm$0.01 \\

\bottomrule
\end{tabular}

\end{table*}

\begin{table*}[h]

\centering
\caption{Ablation results for the movie environment using  the following settings: LLM \textit{Mistral 7B}, 0-9 rating scale, 2-shot, custom system prompt, and perturbation \textit{none}. For each prompt component we show the aggregated score and specific sub-scores for the various test cases.}
\label{table:ablations_movies_retrival_full}
\begin{tabular}{m{0.25\textwidth}
m{0.12\textwidth}
m{0.13\textwidth}
m{0.12\textwidth}
m{0.10\textwidth}
m{0.11\textwidth}}
\toprule
Retrieval \newline component & Genres $\uparrow$ & High/Low $\uparrow$  & Collection \newline of movies $\uparrow$ & Similarity \newline to ML $\uparrow$ & Agg. score $\uparrow$  \\
\midrule
Features similarity & 0.79$\pm$0.01 & \textbf{1.00$\pm$0.00} &  \textbf{0.68$\pm$0.01} & \textbf{0.78$\pm$0.00} & \textbf{0.81$\pm$0.00} \\
T5 similarity & 0.79$\pm$0.00 & \textbf{1.00$\pm$0.00} & 0.67$\pm$0.03 & \textbf{0.78$\pm$0.00} & \textbf{0.81$\pm$0.01} \\
Most recent &  \textbf{0.80$\pm$0.00} & \textbf{1.00$\pm$0.00} & 0.62$\pm$0.01 & \textbf{0.78$\pm$0.00} & 0.80$\pm$0.00 \\
None & \textbf{0.80$\pm$0.00} & \textbf{1.00$\pm$0.00} & 0.49$\pm$0.01 & \textbf{0.78$\pm$0.00} & 0.77$\pm$0.00 \\
\bottomrule
\end{tabular}

\end{table*}

\begin{table*}[h]
\caption{Ablation results for the movie environment using the following settings: LLM \textit{Mistral 7B}, 0-9 rating scale, 2-shot, custom system prompt, \textit{T5-similarity} movie retrieval. For each prompt component we show the aggregated score and specific sub-scores for the various test cases.}
\label{table:ablations_movies_perturbator_full}
\begin{tabular}{m{0.25\textwidth}
m{0.12\textwidth}
m{0.13\textwidth}
m{0.12\textwidth}
m{0.10\textwidth}
m{0.11\textwidth}}
\toprule
Perturbator \newline component & Genres $\uparrow$ & High/Low $\uparrow$  & Collection \newline of movies $\uparrow$ & Similarity \newline to ML $\uparrow$ & Agg. score $\uparrow$  \\
\midrule
gaussian  & \textbf{0.79$\pm$0.00} & \textbf{1.00$\pm$0.00} & \textbf{0.68$\pm$0.01} & \textbf{0.82$\pm$0.00} & \textbf{0.82$\pm$0.00} \\
greedy & 0.78$\pm$0.01 & \textbf{1.00$\pm$0.00} & 0.67$\pm$0.01 & 0.81$\pm$0.00 & \textbf{0.82$\pm$0.00} \\
none & \textbf{0.79$\pm$0.00} & \textbf{1.00$\pm$0.00} & 0.67$\pm$0.03 & 0.78$\pm$0.00 & 0.81$\pm$0.01 \\ 
\bottomrule
\end{tabular}

\end{table*}

\begin{figure*}[h]
    \centering
    \includegraphics{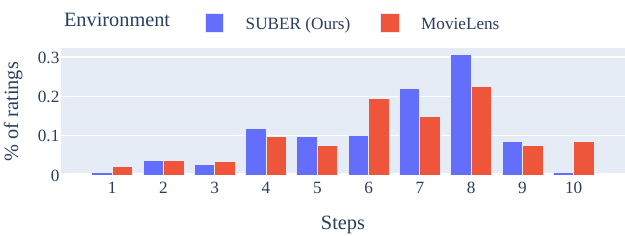}
    \caption{Rating distribution for SUBER movie environment is shown in blue, while the distribution for MovieLens is displayed in red.}
    \vspace{2em}
    \label{fig:enter-label}
\end{figure*}

\begin{table*}[t]
\centering
\caption{Ablation results for the movie environment using \textit{Vicuna-v1.5-13B} as our environment. We test the LLM on coherency and realistic ratings for user-movie interactions. We achieve best performance with 0-9 digit rating scale, 2-shot prompting, and our custom system prompt.}
\label{table:ablations_movies_prompt_vicuna}
\begin{tabular}{
m{0.08\linewidth}
m{0.07\linewidth}
m{0.07\linewidth}|
m{0.12\textwidth}
m{0.13\textwidth}
m{0.12\textwidth}
m{0.10\textwidth}
m{0.11\textwidth}}
\toprule
\multicolumn{3}{c}{Prompt component} \\
Rating scale  & N-shot & System prompt & Genres $\uparrow$ & High/Low $\uparrow$  & Collection \newline of movies $\uparrow$ & Similarity \newline to ML $\uparrow$ & Agg. score $\uparrow$  \\
\midrule
0-9 & 0-shot & default & 0.65$\pm$0.00 & 0.99$\pm$0.00 & 0.62$\pm$0.02 & 0.64$\pm$0.00 & 0.72$\pm$0.00 \\

0-9 & 0-shot& custom & 0.69$\pm$0.00 & 0.99$\pm$0.00 & 0.64$\pm$0.02 & 0.65$\pm$0.00 & 0.74$\pm$0.00 \\

0-9 & 1-shot & default & 0.61$\pm$0.00 & \textbf{1.00$\pm$0.00} & 0.71$\pm$0.01 & \textbf{0.75$\pm$0.00}  & 0.77$\pm$0.00 \\
0-9 & 1-shot & custom & \textbf{0.72$\pm$0.00} & \textbf{1.00$\pm$0.00} & 0.74$\pm$0.03 & 0.74$\pm$0.00 & 0.80$\pm$0.01 \\
0-9 & 2-shot & default & 0.63$\pm$0.01 & \textbf{1.00$\pm$0.00} & 0.81$\pm$0.02 & 0.74$\pm$0.00 & 0.80$\pm$0.00 \\
0-9 & 2-shot  & custom & 0.69$\pm$0.00 & \textbf{1.00$\pm$0.00} & \textbf{0.82$\pm$0.02} & \textbf{0.75$\pm$0.00} & \textbf{0.81$\pm$0.00} \\

1-10 & 2-shot & custom & 0.64$\pm$0.01 & 0.72$\pm$0.03 & 0.68$\pm$0.01 & 0.72$\pm$0.00 & 0.69$\pm$0.01 \\
one-ten & 2-shot & custom & 0.71$\pm$0.01 & \textbf{1.00$\pm$0.00} & 0.72$\pm$0.03 & 0.64$\pm$0.00 & 0.77$\pm$0.01 \\

\bottomrule
\end{tabular}

\end{table*}

\label{app:ablations_books_full_results}
\begin{table*}[h]
\centering
\caption{Ablation results for the book environment using the following settings: 1-5 rating scale, 2-shot, custom system prompt, \textit{T5-similarity} book retrieval, and \textit{no perturbation}. For each prompt component we show the aggregated score and specific sub-scores for the various test cases.}
\label{table:ablations_books_models_full}
\begin{tabular}{m{0.15\textwidth}m{0.07\textwidth}
m{0.12\textwidth}
m{0.13\textwidth}
m{0.12\textwidth}
m{0.11\textwidth}}
\toprule
LLM & Size & Category $\uparrow$ & High/low  $\uparrow$ & Collection \newline of books $\uparrow$ & Agg. score  $\uparrow$ \\
\midrule
GPT-4 & 1760B  & \textbf{0.96$\pm$0.00} & \textbf{1.00$\pm$0.00} & \textbf{0.97$\pm$0.00} & \textbf{0.98$\pm$0.00} \\
GPT-3.5 & 175B & 0.65$\pm$0.00 & 0.99$\pm$0.00 & 0.63$\pm$0.00 & 0.76$\pm$0.00 \\
\midrule
 \multirow{3}{*}{Llama-2-Chat}  & 70B & 0.95$\pm$0.00 &  \textbf{1.00$\pm$0.00} & 0.74$\pm$0.02 & 0.90$\pm$0.01 \\
 & 13B & 0.76$\pm$0.00 &  \textbf{1.00$\pm$0.00} & 0.78$\pm$0.06 & 0.85$\pm$0.02 \\
 & 7B & 0.65$\pm$0.01 &  \textbf{1.00$\pm$0.00} & 0.73$\pm$0.02 & 0.79$\pm$0.01 \\
 \midrule
 \multirow{3}{*}{Vicuna-v1.3} 
 & 33B & 0.63$\pm$0.00 & 0.95$\pm$0.00 & 0.71$\pm$0.00 & 0.76$\pm$0.00 \\
 & 13B & 0.51$\pm$0.02 & 0.87$\pm$0.02 & 0.58$\pm$0.01 & 0.66$\pm$0.02 \\
 & 7B & 0.45$\pm$0.03 & 0.74$\pm$0.01 & 0.57$\pm$0.02 & 0.60$\pm$0.02 \\
 \midrule
 \multirow{3}{*}{Vicuna-v1.5} & 13B & 0.75$\pm$0.01 & 0.99$\pm$0.00 & 0.83$\pm$0.03 & 0.86$\pm$0.01 \\
 & 7B & 0.56$\pm$0.01 & 0.66$\pm$0.02 & 0.72$\pm$0.02 & 0.64$\pm$0.01 \\
 \midrule
\multirow{3}{*}{Mistral} & 8x7B & 0.92$\pm$0.00 &  \textbf{1.00$\pm$0.00} & 0.89$\pm$0.01 & 0.94$\pm$0.00 \\
& 7B & 0.85$\pm$0.00 &  \textbf{1.00$\pm$0.00} & 0.76$\pm$0.02 & 0.87$\pm$0.01 \\
\bottomrule
\end{tabular}

\end{table*}

\begin{table*}[h]
\centering
\caption{Ablation results for the book environment using the following settings: LLM \textit{Mistral 7B} 1-5 rating scale, 2-shot, custom system prompt, and \textit{no perturbation}. For each prompt component we show the aggregated score and specific sub-scores for the various test cases.}
\label{table:ablations_books_retrival_full}
\begin{tabular}{m{0.25\textwidth}
m{0.12\textwidth}
m{0.13\textwidth}
m{0.12\textwidth}
m{0.11\textwidth}}
\toprule
Retrieval \newline component & Category $\uparrow$ & High/low  $\uparrow$ & Collection \newline of books $\uparrow$ & Agg. score $\uparrow$ \\
\midrule
None & 0.85$\pm$0.00 & 0.99$\pm$0.00 & 0.50$\pm$0.00 & 0.78$\pm$0.00 \\
Most recent & \textbf{0.86$\pm$0.00} &  \textbf{1.00$\pm$0.00} & 0.64$\pm$0.01 & 0.83$\pm$0.00 \\
T5 similarity  & 0.85$\pm$0.00 &  \textbf{1.00$\pm$0.00} & 0.76$\pm$0.02 & 0.87$\pm$0.01 \\
Features similarity & 0.85$\pm$0.00 &  \textbf{1.00$\pm$0.00} & \textbf{0.79$\pm$0.02} & \textbf{0.88$\pm$0.01} \\
\bottomrule
\end{tabular}

\end{table*}

\begin{table*}[h]
\centering
\caption{Ablation results for the book environment using the following settings: LLM \textit{Mistral 7B} 1-5 rating scale, 2-shot, custom system prompt, \textit{T5-similarity} book retrieval. For each prompt component we show the aggregated score and specific sub-scores for the various test cases.}
\label{table:ablations_books_perturbator_full}
\begin{tabular}{m{0.25\textwidth}
m{0.12\textwidth}
m{0.13\textwidth}
m{0.12\textwidth}
m{0.11\textwidth}}
\toprule
Perturbator \newline component & Category $\uparrow$ & High/low  $\uparrow$ & Collection \newline of books $\uparrow$ & Agg. score $\uparrow$ \\
\midrule
None & \textbf{0.85$\pm$0.00} &  \textbf{1.00$\pm$0.00} & \textbf{0.76$\pm$0.02} & \textbf{0.87$\pm$0.01} \\
Greedy & 0.82$\pm$0.00 & 0.95$\pm$0.00 & \textbf{0.76$\pm$0.05} & 0.84$\pm$0.02 \\
Gaussian & 0.81$\pm$0.00 & 0.91$\pm$0.00 & 0.75$\pm$0.02 & 0.83$\pm$0.01 \\
\bottomrule
\end{tabular}

\end{table*}

\FloatBarrier
\clearpage

\section{Details Human Evaluation}
\label{app:user_study}

\begin{lstlisting}[style=interaction, label=listing:user_study_example, caption={ Example LLM interaction of the human evaluation }]
<template>Question:
Samuel is a 24 years old man, he is a student who loves chemistry and watch movies on his free time. His favorite genres are drama, comedy, action and crime. He enjoys watching movies that make him think and feel, and that have a good plot and good acting. He likes to watch movies that are based on true events or that have a good message. He dislikes science fiction, horror and documentary movies, as they are too unrealistic or too intense for him. He likes to watch movies with his friends and classmates, and he always looks for a good laugh or a good cry. Samuel is a very analytical person, who loves to read and research about different topics. He is very focused on his studies, and he is always looking for new challenges. He is a good listener and a good conversationalist, and he likes to hear different perspectives.
Samuel has previously watched the following movies (in parentheses are the ratings he gave on a scale of 1 to 10): "Blade Runner 2049" (5), "Inception" (10), "A Beautiful Mind" (9).
Consider the movie "E.T. the Extra-Terrestrial", released in 1982, which is described as follows: an alien is left behind on Earth and saved by the 10-year-old Elliot who decides to keep him hidden in his home. While a task force hunts for the extra-terrestrial, Elliot, his brother, and his little sister Gertie form an emotional bond with their new friend, and try to help him find his way home. The movie "E.T. the Extra-Terrestrial" contains the following genres:
-science fiction
-adventure
-family
-fantasy
Here are the 2 main actors of the movie, in order of importance: Henry Thomas (M), Drew Barrymore (F). On average, people rate the movie "E.T. the Extra-Terrestrial" 7.5 on a scale of 1 to 10. Samuel watches the movie "E.T. the Extra-Terrestrial" for the 1st time.
What can you conclude about Samuel's rating for the movie "E.T. the Extra-Terrestrial" on a scale of 1 to 10, where 1 represents a low rating and 10 represents a high rating, based on available information and logical reasoning?

Answer:
Based on Samuel's preferences and tastes, I conclude that he will assign a rating of 8 on a scale of 1 to 10, because 
1) He enjoys watching movies that make him think and feel, and have good plots and acting, which is evident from his previous ratings of "A Beautiful Mind" (9) and "Inception" (10); 
2) "E.T. the Extra-Terrestrial" has elements of adventure and family, which align with his preferred genres; and 
3) Although it belongs to the science fiction genre, which Samuel generally dislikes, the film's emotional themes and focus on friendship could resonate well enough with him to warrant a higher rating.
</template>
\end{lstlisting}

In the human evaluation, we show participants pairs of prompt and LLM response as shown in  \Cref{listing:user_study_example} and ask them to assign a score between 1 and 5 according to the following guideline:
\begin{enumerate}
    \item The LLM's answer does not make sense.
    \item The rating and explanation do not match the person's interests and preferences.
    \item The answer makes some sense, but a different rating would be more appropriate.
    \item The rating and explanation make sense, but not all important information is considered.
    \item The rating and explanation perfectly match the user's interests and you completely agree with the rating.
\end{enumerate}

We selected 10 different queries, 4 from the genres test set of the ablation study, 2 from the movie franchise test set of the ablation study, and the remaining 4 we randomly sampled user and movie while manually filling in the previous ratings of the selected users. We successively queried each model with all ten questions using the default configuration (2 shot prompt and custom system prompt with no perturbator), this way we collected 50 different answers, ten for each of the five models.
We perform the study by asking 14 participants to score the LLM interactions. 
In the questionnaire, the order of the questions is randomized for each participants. We then aggregate the score by averaging over each model.

\clearpage

\section{RL Models}\label{app:rl_model_training}
For A2C and PPO we implemented the actor based on the principles of low-rank approximation \citep{aggarwal2016recommender}. For each user $u$ within the set $U$, we maintain a feature vector $e_u$. Similarly, for each movie $m$, we use its feature vector $e_m$ and bias $b_m$. Additionally, we introduce the movie embedding matrix $E$, and the bias vector $b$. The probability of recommending movie $m$ to user $u$ is calculated as follows:
\begin{equation}
   \label{eq:sampleA2C}
    \text{softmax}\left(A + E\cdot e_u + b\right)_m,
\end{equation}
where $A$ serves as a mask to assign a probability of zero to movies that user $u$ has already viewed. In other words, the entry $A_m$ is set to negative infinity if user $u$ has previously watched movie $m$. We employ A2C~\citep{mnih2016asynchronous} to train the agent. The actor network, which is responsible for recommending movies, samples actions according to \Cref{eq:sampleA2C}, while the critic consists of a basic two-layer neural network, which takes the user together with the past movie ratings of users as input. We train the model with the default configuration of SB3 \citep{stable-baselines3} for 1.6M steps on SUBER. All parameters are default, except for gamma, which is changed to 0.975.
For TRPO and DQN the actor network additionally takes as input the past movie ratings. Also in this case, the models were trained for 1.6M steps with the default configuration of SB3, and all parameters are default except for gamma, which is changed to 0.975.

\subsection{Metrics}
In this section we outline the different metrics used for the RL model evaluation. In different metrics we need to distinguish if an item is relevant or not, since we don't have binary interaction, we consider an item relevant for a specific user if the rating given by the user is seven or above. This choice is motivated by \citet{ramos2015statistical}.

\shortp{Mean Average Precision (MAP@10)}

Let $\text{Precision@}k$ be the proportion of relevant items in the top-k recommendations for a fixed user, i.e. 
$$\text{Precision@}k = \frac{\#\text{relevant items in the first }k \text{ recommendations}}{k}.$$ 
For a fixed user, the Average Precision (AP@10) is defined as the average of the precision values calculated at the positions where relevant items appear within the top ten recommendations. 
$$AP@10 = \frac{1}{10} \sum_{k=1}^{10} \text{Precision@}k\cdot R(k), $$ where R(k) is 1 if item k is relevant and 0 otherwise. MAP@10 is the average of AP@10 across users.

MRR@$10$ combines precision and relevance across multiple users to provide a comprehensive measure of how well the system ranks relevant items within the top-k recommendations.

\shortp{Mean Reciprocal Rank (MRR@10)}
Mean Reciprocal Rank is used to understand how many recommendations are needed on average before the first relevant item is recommended, and thus measure how quickly a user is satisfied.

For a fixed user, the Reciprocal Rank (RR) is the multiplicative inverse of the rank of the first relevant item. RR@10 is the RR when considering the first ten recommendations (if no relevant item is recommended in the first 10 recommendations, then the value is set to 0). MRR@10 is the average of RR@10 across users.

\shortp{Personalization (Pers.@10)}
To assesses whether a model recommends many of the same items to different users, we use Personalization Pers.@$10$, which is defined as (1 - cosine similarity) between user's lists of top 10 recommendations.

% Define the user-item interaction matrix
Let $ M $ be the user-item interaction matrix, where $ M_{u,i} = 1 $ if user $ u $ has been recommended item $i$ in the first 10 recommendations and $ M_{u,i} = 0 $ otherwise.

% Define the cosine similarity between two users
The cosine similarity $ \text{cos\_sim}(u, v) $ between users $ u $ and $ v $ is defined as:
$$
\text{cos\_sim}(u, v) = \frac{\sum_{i} M_{u,i} M_{v,i}}{\sqrt{\sum_{i} M_{u,i}^2} \sqrt{\sum_{i} M_{v,i}^2}}
$$

% Define the set of user pairs
Let $ \mathcal{P} $ be the set of all user pairs $ (u, v) $ such that $ u > v $.

% Compute the personalization metric
The personalization metric is defined as:
$$
\text{Pers.}@10 = 1 - \frac{1}{|\mathcal{P}|} \sum_{(u, v) \in \mathcal{P}} \text{cos\_sim}(u, v)
$$

A high personalization score indicates that recommendations vary significantly among different users, suggesting that the model offers a more personalized experience to each user.

\shortp{Liked Genres}
Each user in the training dataset has both preferred and disliked movie genres. The trained RL model generates a list of the top 5 movie recommendations for each user. The recommendations are classified into three categories: liked (movies matching preferred genres and excluding disliked ones), disliked (movies with disliked genres and no preferred ones), and neutral (remaining recommendations). ``Liked Genres'' is the percentage of recommendations that fall into the first category.
For each user $u$ we define $L(u)$ to be the percentage of movies that fall into the liked genre category in the top-5 recommendation, then the liked genre metric is the average over the dataset of users:
$$
\text{Linked Genres} = \frac{1}{\vert U \vert}\sum_{u \in U}L(u).
$$
This metric measures how well the recommender system makes genre-specific recommendations.

\subsection{Genre Preference}
\begin{figure}[ht]
    \centering
     \includegraphics[width=0.7\linewidth]{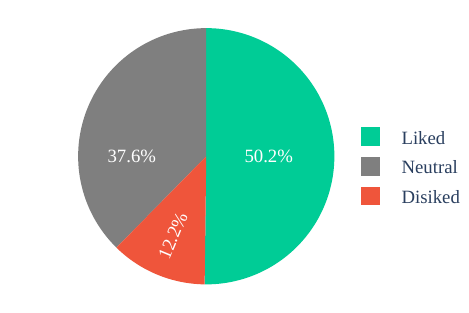}
    \caption{User genre preference statistic of top-5 movie recommendations generated by the RL model, trained with A2C. }
    \label{fig:genres_recommended_pie_A2C}
\end{figure}

\begin{figure}[ht]
    \centering
     \includegraphics[width=0.7\linewidth]{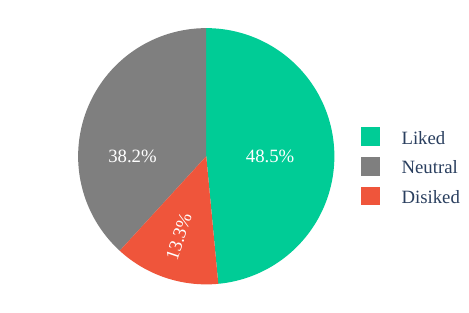}
    \caption{User genre preference statistic of top-5 movie recommendations generated by the RL model, trained with TRPO.}
    \label{fig:genres_recommended_pie_TRPO}
\end{figure}

\begin{figure}[ht]
    \centering
     \includegraphics[width=0.7\linewidth]{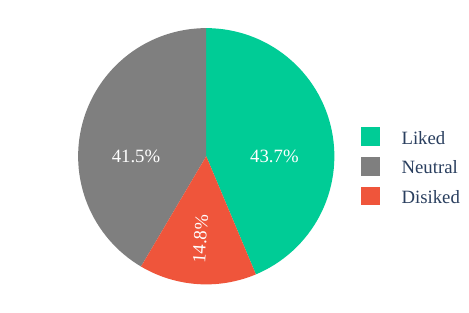}
    \caption{User genre preference statistic of top-5 movie recommendations generated by the RL model, trained with PPO.}
    \label{fig:genres_recommended_pie_PPO}
\end{figure}

\begin{figure}[h]
    \centering
     \includegraphics[width=0.7\linewidth]{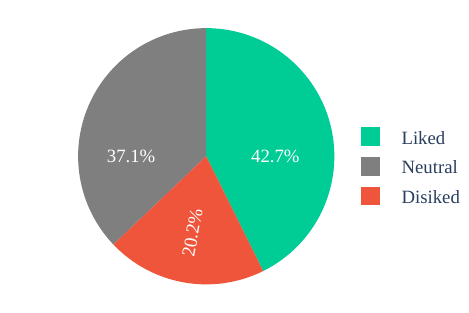}
    \caption{User genre preference statistic of top-5 movie recommendations generated by the RL model, trained with DQN.}
    \label{fig:genres_recommended_pie_DQN}
\end{figure}

In this section, we present the user-genre preference statistics for the top-5 recommendations generated by A2C, TRPO, PPO, and DQN models. The results show that, for all models, the majority of recommendations fall into the \textit{liked} category. By comparing \Cref{fig:reward_during_training} with \Cref{fig:genres_recommended_pie_A2C,fig:genres_recommended_pie_TRPO,fig:genres_recommended_pie_PPO,fig:genres_recommended_pie_DQN}, we observe a positive correlation between model rewards and the percentage of top-5 recommended movies in the "liked" category. In other words, the better the model is at recommending movies that align with users' preferred genres, the higher the reward it achieves in the environment.

% We observe from \Cref{fig:reward_during_training} how A2C is the model that learns better our environment. 
Moreover, as shown in \Cref{fig:genres_recommended_pie_A2C,fig:genres_recommended_pie_TRPO,fig:genres_recommended_pie_PPO,fig:genres_recommended_pie_DQN}  the RL recommender model is able to learn the dynamics of genre preferences of users, mainly recommending movies that fall into the favored genres of users. It should be noted that recommending a \textit{neutral} movie can be a valid strategy, especially if it is a highly praised or outstanding movie.


\clearpage
\subsection{Recommended Movies Examples}
For a set of random users, we interact with the trained RL model, in \Cref{tab:rec_max_5,tab:rec_ava_5,tab:rec_maya_5} with embedding dim 32, and show the first 5 recommended movies. 

\begin{table*}[ht]
    \centering
      \caption{At the top we show the personal interest of Max. At the bottom, we show the title and genres of the first 5 recommended movies. }
    \label{tab:rec_max_5}
    \begin{tabular}{m{0.45\textwidth}m{0.45\textwidth}}
    \toprule
    \textbf{Name} & Max \\
    \textbf{Liked genres} & thriller, documentary, fantasy, crime \\
    \textbf{Disliked genres} & romance \\
    \midrule
     \textbf{Recommended movie} & \textbf{Genres} \\
     \midrule
     The Walk & adventure, drama, thriller, history\\
     The Testament of Dr. Mabuse & crime, mystery, thriller \\
     The Island	 &	action, thriller, science fiction, adventure \\ 
     Captain America &	action, adventure, science fiction, war \\ 	
     The Count of Monte Cristo &	adventure, drama, history \\
    \bottomrule
    \end{tabular}
  
\end{table*}
\vspace{-1em}

\begin{table*}[ht]
    \centering
    \caption{At the top we show the personal interest of Ava. At the bottom, we show the title and genres of the first 5 recommended movies. }
    \label{tab:rec_ava_5}
    \begin{tabular}{m{0.45\textwidth}m{0.45\textwidth}}
    \toprule
    \textbf{Name} & Ava \\
    \textbf{Liked genres} & drama, science fiction, animation, adventure \\
    \textbf{Disliked genres} & romance, fantasy, crime, comedy \\
    \midrule
     \textbf{Recommended movie} & \textbf{Genres} \\
     \midrule
     Inception & action, science fiction, adventure \\
     The Walk & adventure, drama, thriller, history\\
     Spartacus & history, war, drama, adventure\\
     Exodus: Gods and Kings & adventure, drama, action \\
     Cliffhanger & action, adventure, thriller \\
     Atragon & science fiction, action, adventure, fantasy \\
    \bottomrule
    \end{tabular}
  
\end{table*}
\vspace{-1em}
\begin{table*}[ht]
    \centering
     \caption{At the top we show the personal interest of Maya. At the bottom, we show the title and genres of the first 5 recommended movies. }
    \label{tab:rec_maya_5}
    \begin{tabular}{m{0.45\textwidth}m{0.45\textwidth}}
    \toprule
    \textbf{Name} & Maya \\
    \textbf{Liked genres} & drama, science fiction, documentary, comedy \\
    \textbf{Disliked genres} & horror, animation \\
    \midrule
     \textbf{Recommended movie} & \textbf{Genres} \\
     \midrule
        The One-Armed Swordsman & action, drama \\
        The One I Love	 & romance, comedy, drama\\
        In the Heart of the Sea	 & thriller, drama, adventure, action, history\\
        Detachment & drama\\
        Amadeus & history, music, drama\\
        Margaret Cho: I'm the One That I Want & comedy \\
    \bottomrule
    \end{tabular}
    
\end{table*}

\vspace{1em}
\subsection{Environment Performance}

\begin{table*}[b]
\vspace{-2em}
       \caption{List of Large Language Models tested on the environment. iterations/seconds are computed for all models using GPTQ and Exllama on a RTX3090, and A100-40GB for Llama-2-70B.}
    \label{tab:llm}
   
    \centering
    \begin{tabular}{cccc}
   \toprule
         Model name & Size & Contex length & iterations/s (in our env) \\
         \midrule
         GPT-4 & 1760B & 8k / 32k & \multirow{2}{*}{ API ratelimit dependent }\\
         GPT-3.5 & 175B  & 4k / 16k & \\
         \midrule
         \multirow{3}{*}{Llama-2-Chat} & 70B  & \multirow{3}{*}{4,096}   & 1.6 \\
         & 13B   &  & 5 \\
         & 7B  &  & 6 \\
         \midrule
         \multirow{3}{*}{Vicuna-v1.3} & 33B  & \multirow{3}{*}{2,048}  & 3 \\
         & 13B &  & 5 \\
         & 7B  &  & 6 \\
         \midrule
         \multirow{2}{*}{Vicuna-v1.5} & 13B  & \multirow{2}{*}{4,096}  & 5 \\
         & 7B  &  & 6 \\
         \midrule
         \multirow{2}{*}{Mistral} & 8x7B  & \multirow{2}{*}{4,096}  & 2 \\
         & 7B  &  & 6 \\
         \bottomrule
    \end{tabular}
 
\end{table*}

\clearpage
\section{List of Hobbies and Professions}
\label{app:list}

\begin{table}

\caption{List of children's hobbies}
\label{tab:hobby_children}
\begin{tabular}{l}
\toprule
Hobby-name \\
\midrule
Drawing and painting \\
Playing piano \\
Playing guitar \\
Playing violin \\
Playing flute \\
Playing drums \\
Dancing \\
Reading books \\
Writing stories \\
Board games  \\
Card games \\
Gardening \\
Cooking \\
Backing \\
Building with Lego \\
Collecting stamps \\
Collecting coins \\
Collecting cards \\
Photography \\
Learning magic tricks \\
Soccer \\
Basketball \\
Swimming \\
Volleyball \\
Tennis \\
Acting \\
Singing \\
Puppetry \\
Birdwatching or nature exploration \\
Science experiments \\
Playing video games \\
Origami \\
Learning a new language \\
\bottomrule
\\
\\
\\
\\
\\
\\
\\
\\
\\
\\
\\
\\
\\
\\
\\
\\
\\
\\
\\
\\
\\
\\
\\
\\
\\
\\
\\
\\
\\
\\
\end{tabular}
 
\end{table}

\begin{table}
\caption{Jobs list}
\begin{tabular}{l}
\toprule
Jobs \\
\midrule
Account Manager \\
Accountant \\
Actor \\
Actuary \\
Administrator \\
Advertising Executive \\
Aerospace Engineer \\
Aerospace Technician \\
Air Traffic Controller \\
Animal Trainer \\
Architect \\
Archivist \\
Art Director \\
Artist \\
Auctioneer \\
Auto Mechanic \\
Baggage Handler \\
Bailiff \\
Baker \\
Banker \\
Barber \\
Barber Shop Owner \\
Barista \\
Bartender \\
Benefits Administrator \\
Bicycle Mechanic \\
Biologist \\
Blacksmith \\
Boat Captain \\
Bodyguard \\
Bookkeeper \\
Botanical Illustrator \\
Botanist \\
Brewery Worker \\
Bricklayer \\
Broadcast Technician \\
Building Inspector \\
Bus Driver \\
Bus Mechanic \\
Butcher \\
CIO (Chief Information Officer) \\
Cabin Crew \\
Cake Decorator \\
Call Center Operator \\
Car Salesperson \\
Carpenter \\
Cartographer \\
Cashier \\
Casino Dealer \\
Caterer \\
Chaplain \\
Chauffeur \\
Chef \\
Chemical Engineer \\
Chemist \\
Chief Financial Officer (CFO) \\
Chimney Sweep \\
Chiropractor \\
Civil Engineer \\
Claims Adjuster \\
Cleaner \\
Clown \\
Coach \\
Coachbuilder \\
Commercial Pilot \\
Composer \\
Computer Programmer \\
Computer Systems Analyst \\
Concierge \\
Conservationist \\

\end{tabular}

\end{table}

\begin{table}
\caption{Jobs list}
\begin{tabular}{l}
Construction Worker \\
Cost Estimator \\
Counselor \\
Courier \\
Court Reporter \\
Craftsperson \\
Cruise Ship Captain \\
Cryptographer \\
Curator \\
Customer Service Representative \\
Dairy Farmer \\
Dancer \\
Data Analyst \\
Data Entry Operator \\
Database Administrator \\
Demolition Worker \\
Dental Hygienist \\
Dentist \\
Designer \\
Desktop Publisher \\
Detective \\
Detective Inspector \\
Dialysis Technician \\
Diesel Mechanic \\
Dietician \\
Digital Marketer \\
Dispatch Operator \\
Doctor \\
Dog Trainer \\
Door-to-Door Salesperson \\
Dressmaker \\
Drummer \\
Dry Cleaner \\
Economist \\
Economist \\
Electrician \\
Engineer \\
Event Planner \\
Farmer \\
Fashion Designer \\
Firefighter \\
Flight Attendant \\
Florist \\
Forensic Scientist \\
Gardener \\
Geologist \\
Graphic Designer \\
Hairdresser \\
Historian \\
Hotel Manager \\
Human Resources Manager \\
Illustrator \\
Industrial Designer \\
Insurance Agent \\
Interior Designer \\
Interpreter \\
Janitor \\
Journalist \\

Judge \\
Laboratory Technician \\
Lawyer \\
Librarian \\
Lifeguard \\
Linguist \\
Makeup Artist \\
\end{tabular}

\end{table}

\begin{table}
\vspace{-1em}
\caption{Jobs list}
\begin{tabular}{l}

Locksmith \\
Manager \\
Marketing Specialist \\
Massage Therapist \\
Mechanic \\
Medical Assistant \\
Meteorologist \\
Model \\
Musician \\
Nanny \\
Nurse \\
Nutritionist \\
Occupational Therapist \\
Optician \\
Painter \\
Paramedic \\
Pharmacist \\
Photographer \\
Physical Therapist \\
Physician Assistant \\
Pilot \\
Plumber \\
Police Officer \\
Politician \\
Postal Worker \\
Producer \\
Professor \\
Psychologist \\
Public Relations Specialist \\
Real Estate Agent \\
Receptionist \\
Reporter \\
Research Scientist \\
Sales Representative \\
Scientist \\
Security Guard \\
Singer \\
Social Media Manager \\
Social Worker \\
Software Developer \\
Sound Engineer \\
Speech Therapist \\
Sports Coach \\
Statistician \\
Stockbroker \\
Surveyor \\
Tailor \\
Teacher \\
Technical Writer \\
Technician \\
Therapist \\
Tour Guide \\
Translator \\
Travel Agent \\
Truck Driver \\
UI/UX Designer \\
Veterinarian \\
Video Editor \\
Waiter/Waitress \\
Web Developer \\
Welder \\
Writer \\
Yoga Instructor \\
Zookeeper \\
\bottomrule
\end{tabular}

\end{table}

%%%%%%%%%%%%%%%%%%%%%%%%%%%%%%%%%%%%%%%%%%%%%%%%%%%%%%%%%%%%%%%%%%%%%%